\documentclass[12pt, draftclsnofoot, onecolumn]{IEEEtran}

\usepackage{amsmath}
\usepackage{amssymb}
\usepackage{amsthm}
\usepackage{algorithm}
\usepackage{algorithmic}
\usepackage{bm}
\usepackage{bbding}
\usepackage{booktabs}
\usepackage{clrscode}
\usepackage{cases}
\usepackage{cite}
\usepackage[dvips]{graphicx}
\usepackage{color}
\usepackage[mathscr]{euscript}
\usepackage{epsfig}
\usepackage{extarrows}
\usepackage{fancyhdr}
\usepackage{hyperref}
\usepackage{lipsum}
\usepackage{lineno} 
\usepackage{microtype}
\usepackage{subfigure}
\usepackage{stfloats}
\usepackage{setspace}
\usepackage{titletoc}
\usepackage{verbatim}

\newcommand{\td}{\hspace{0.02cm}\mathsf{td}}
\newcommand{\hy}{\hspace{0.02cm}\mathsf{hy}}
\newcommand{\no}{\hspace{0.02cm}\mathsf{no}}
\newcommand{\bv}{{\bf{v}}}
\newcommand{\tv}{{\bf{v}}}

\newcommand{\teq}{\triangleq}
\newcommand{\mbf}{\mathbf}
\newcommand{\mbb}{\mathbb}
\newcommand{\mcal}{\mathcal}
\newcommand{\tbf}{\textbf}

\newcommand{\dia}{\text{diag}}
\newcommand{\su}{\text{sum}}

\newcommand{\bs}{\boldsymbol}
\newcommand{\IRS}{\bm{\Phi}}
\newcommand{\bh}{\bs{h}}

\newcommand{\noi}{\sigma^2}

\newcommand{\m}{\hspace{0.02cm} \mathsf{m}}

\newcommand{\bb}{\bs{b}}

\newcommand{\bg}{\bs{g}}

\newenvironment{myproof}{{\it Proof:}}{\hfill$\blacksquare$}
\theoremstyle{theorem}
\newtheorem{Pro}{Proposition}
\newtheorem{Lem}{Lemma}
\theoremstyle{definition}
\newtheorem{Rem}{Remark}

\begin{document}
    \linespread{1.4}
    \title{Delay-aware Multiple Access Design for Intelligent Reflecting Surface Aided Uplink Transmission}
    \author{Piao Zeng, Guangji Chen, Qingqing~Wu,~\IEEEmembership{Senior Member,~IEEE}, Deli Qiao, and Abbas~Jamalipour,~\IEEEmembership{Fellow,~IEEE}
    \thanks{P. Zeng is with the School of Communication and Electronic Engineering, East China Normal University, Shanghai, China  (e-mail: 52181214005@stu.ecnu.edu.cn).
    G. Chen is with the State Key Laboratory of Internet of Things for Smart City, University of Macau, Macau, China (email: guangjichen@um.edu.mo).
    Q. Wu is with the Department of Electronic Engineering, Shanghai Jiao Tong University, Shanghai, China (e-mail: qingqingwu@sjtu.edu.cn).
    D. Qiao is with the School of Communication and Electronic Engineering, East China Normal University, Shanghai, China, and also with the National Mobile Communications Research Laboratory, Southeast University, Nanjing, China (e-mail: dlqiao@ce.ecnu.edu.cn).
    Abbas Jamalipour is with the School of Electrical and Information Engineering, The University of Sydney, Sydney, NSW 2006, Australia (e-mail: a.jamalipour@ieee.org). }}
    \maketitle
	
    \begin{abstract}
    In this paper, we develop a {hybrid} multiple access (MA) protocol for an intelligent reflecting {surface} (IRS) aided uplink transmission network {by incorporating the IRS-aided time-division MA (I-TDMA) protocol and the IRS-aided non-orthogonal MA (I-NOMA) protocol as special cases}. Two typical communication scenarios, namely the transmit power limited case and the transmit energy limited case are considered, where the device's rearranged order, time and power allocation, as well as dynamic IRS beamforming patterns over time are jointly optimized to minimize the sum transmission delay. To shed light on the superiority of the proposed IRS-aided hybrid MA (I-HMA) protocol over conventional protocols, the conditions under which I-HMA outperforms I-TDMA and I-NOMA are revealed by characterizing their corresponding optimal solution. Then, a computationally efficient algorithm is proposed to obtain the high-quality solution to the corresponding optimization problems. Simulation results validate our theoretical findings, demonstrate the superiority of the proposed design, and draw some useful insights. Specifically, it is found that the proposed protocol can significantly reduce the sum transmission delay by combining the additional gain of dynamic IRS beamforming with the high spectral efficiency of NOMA, which thus reveals that integrating IRS into the proposed HMA protocol is an effective solution for delay-aware optimization.
    Furthermore, it reveals that the proposed design reduces the time consumption not only from the system-centric view, but also from the device-centric view.
    \end{abstract}
    
    
    \linespread{1.8}
    \section{Introduction}
    \IEEEPARstart{U}{nder} the trends in establishing the Internet-of-Things (IoT) ecosystem, the next generation wireless networks are expected to support an unprecedentedly large number of connections \cite{chettri2020comprehensive}. To accommodate the growing demands of connectivity, the multiple access (MA) technique is one of the most fundamental enablers for facilitating data transmission from massive IoT devices to the access point (AP) or the base station (BS).
    The existing MA techniques can be loosely classified into two categories, namely the orthogonal multiple access (OMA) and non-orthogonal multiple access (NOMA).
    Different from OMA where one resource block (in time, frequency, or code) is occupied by at most one user, NOMA allows different users to share the same time/frequency resources and to be multiplexed in different power levels by invoking superposition coding and successive interference cancellation (SIC) techniques  \cite{dai2015non,liu2017nonorthogonal,ding2017survey,ding2018impact}.
    
    In general, the optimal choice of MA protocols may change in terms of different application scenarios and objectives. For example, for the \tbf{transmit power limited system}, the authors in \cite{chen2017optimization} theoretically compared the optimum sum-rate performance for NOMA and OMA systems with the consideration of user fairness, which revealed that NOMA always achieves better performance than the conventional OMA. Apart from the single-input single-output system as studied in \cite{chen2017optimization}, the authors
    in \cite{zeng2017capacity} investigated the performance of a multiple-input multiple-output NOMA network and analytically proved the superiority of employing NOMA over OMA in terms of both sum channel capacity and ergodic sum capacity. 
    In contrast, for the \tbf{transmit energy limited system}, the authors in \cite{wu2018spectral} proved that the NOMA strategy is neither spectral efficient nor energy efficient when the circuit energy consumption is non-negligible for IoT devices, e.g. in wireless powered communication networks.
    Furthermore, the error propagation of SIC in NOMA may become more severe as the number of users increases \cite{zeng2019energy}. 
    Apart from the clustered hybrid MA scheme proposed in \cite{zeng2019energy}, the authors in \cite{ding2018delay} designed a two-user paired hybrid NOMA scheme to minimize the sum transmission delay for a  mobile edge computing (MEC) system.
    As a generalization of \cite{ding2018delay}, the authors in \cite{zeng2019delay} and \cite{zhu2020resource} extended the two-user scenario to a multi-user one via an iterative procedure and a grouping strategy, respectively. A more general MA strategy for multiple users was developed in \cite{ding2022hybrid} by including pure OMA and pure NOMA strategy as special cases.

    \textls[1]{Although the efficiency of the transmission can be improved with well-designed MA schemes, the throughput of the transmission will degrade severely due to the attenuation of the wireless channels between transceivers, which thus limits the performance of IoT networks. Fortunately, thanks to the significant progress in electrocircuit and metamaterial \cite{kaina2014shaping}, intelligent reflecting surface (IRS) has recently emerged as a promising technology to improve the spectral and energy efficiency for the next generation wireless networks\cite{wu2019intelligent}.
    Specifically, with the assistance of a large number of software-controlled reconfigurable passive elements, the IRS can modify the phase shift and/or the amplitude of the incident signals intelligently so as to create a favorable signal propagation environment \cite{wu2019beamforming}. {Benefiting from the unique physical structure, IRS has the great potential to solve the existing issues in wireless communications \cite{wu2019towards}, which has attracted intensive research interest \cite{wu2021irs,chen2021irs,zheng2020intelligent_lett,zhou2020delay,mu2021joint,bansal2021rate,mei2022multi,mei2022intelligent,al2021reconfigurable,hua2021intelligent,zeng2022delay,shao2022Target,ding2020on,yue2022Performance,chen2022active}.} For example, the authors in \cite{zheng2020intelligent_lett} presented the relationship of minimum transmit powers required by different MA schemes in the IRS-assisted wireless network. Specifically, the results in  \cite{zheng2020intelligent_lett} showed that the performance comparison between NOMA and  time-division MA (TDMA) in the IRS-assisted system depends on the target rates and locations of the users, which is not fully comply with the conclusions in conventional systems without IRS.
  To achieve the delay-optimal users' scheduling for an IRS-aided MEC system, the authors in \cite{zhou2020delay} proposed a flexible time-sharing NOMA scheme, which suggested that different from the case without IRS, the performance gap between the NOMA and TDMA benchmarks with IRS is much smaller. 
    Moreover, there are also works on IRS-aided multi-user communications with other MA strategies, such as rate-splitting MA schemes \cite{bansal2021rate}, beam/angle schemes division MA \cite{mei2022multi,mei2022intelligent}, and sparse code MA schemes \cite{al2021reconfigurable}.}

    Despite of the above works, some fundamental issues still remain to be explored in IRS-aided IoT networks. 
    \tbf{First}, as the transmission delay has become a critical bottleneck that limits the progress of the newly emerged latency-sensitive applications in the IoT networks, such as virtual reality and autonomous driving,
    delay-aware optimization is of great practical significance. However, the related design for the new paradigm of IRS-aided systems is still in its infancy. Thus, this calls for the efforts to establish a more advanced MA protocol to fully reap the potential performance gain of employing the IRS to assist the multi-device uplink (UL) transmission networks. On one hand, under the conventional TDMA protocol, the IRS can achieve additional beamforming (BF) gain by employing multiple IRS BF patterns in a channel coherence interval, i.e., dynamic IRS BF \cite{wu2021irs}, which is shown to outperform NOMA in the aspect of spectral efficiency \cite{chen2021irs}. On the other hand, under the conventional NOMA protocol, the delay of the UL transmission can be significantly reduced compared with the TDMA protocol \cite{ding2018delay}.
    \textls[0]{Considering the above factors, it deserves dedicated work to custom design a proper MA scheme for the IRS-aided system to minimize the sum UL transmission delay. \tbf{Second}, how to enable more flexible resource allocation and IRS BF design to minimize the sum transmission delay of the system? This question is driven by the fact that, unlike conventional transmission networks, the use of IRS provides new degrees of freedom (DoFs) to combat the undesired channel condition and further reduce the transmission delay.
    Nevertheless, it also introduces new optimization variables and makes the IRS BF vectors and transmit power of each device as well as the time allocation closely coupled in the constraints,} thus rendering the joint design of the IRS BF and resource allocation more challenging than that in conventional networks.

\begin{figure}[t]
    \flushright
    \begin{minipage}[t]{0.35\textwidth}
    \flushleft 
    \setlength{\abovecaptionskip}{0.45cm}
        \subfigure{
        \includegraphics[height=0.69\textwidth]{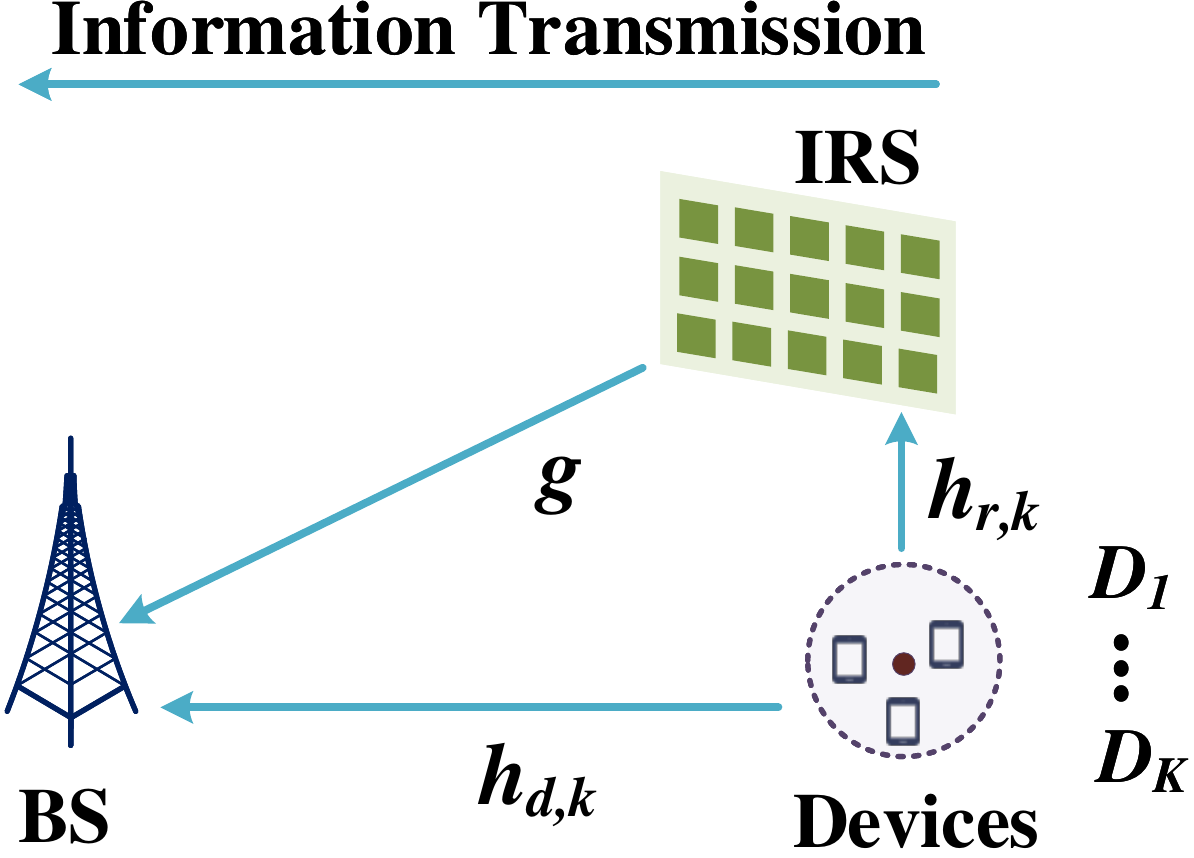}}
        \caption{An IRS-aided UL transmission system.}
        \label{fig:Channel}\vspace{-0.4cm}
    \end{minipage}\hspace{0.5cm}
    \begin{minipage}[t]{0.6\textwidth}
    \flushright
    \subfigure[IRS-aided HMA]{
        \label{fig:HNOMA}
        \includegraphics[height=0.382\textwidth]{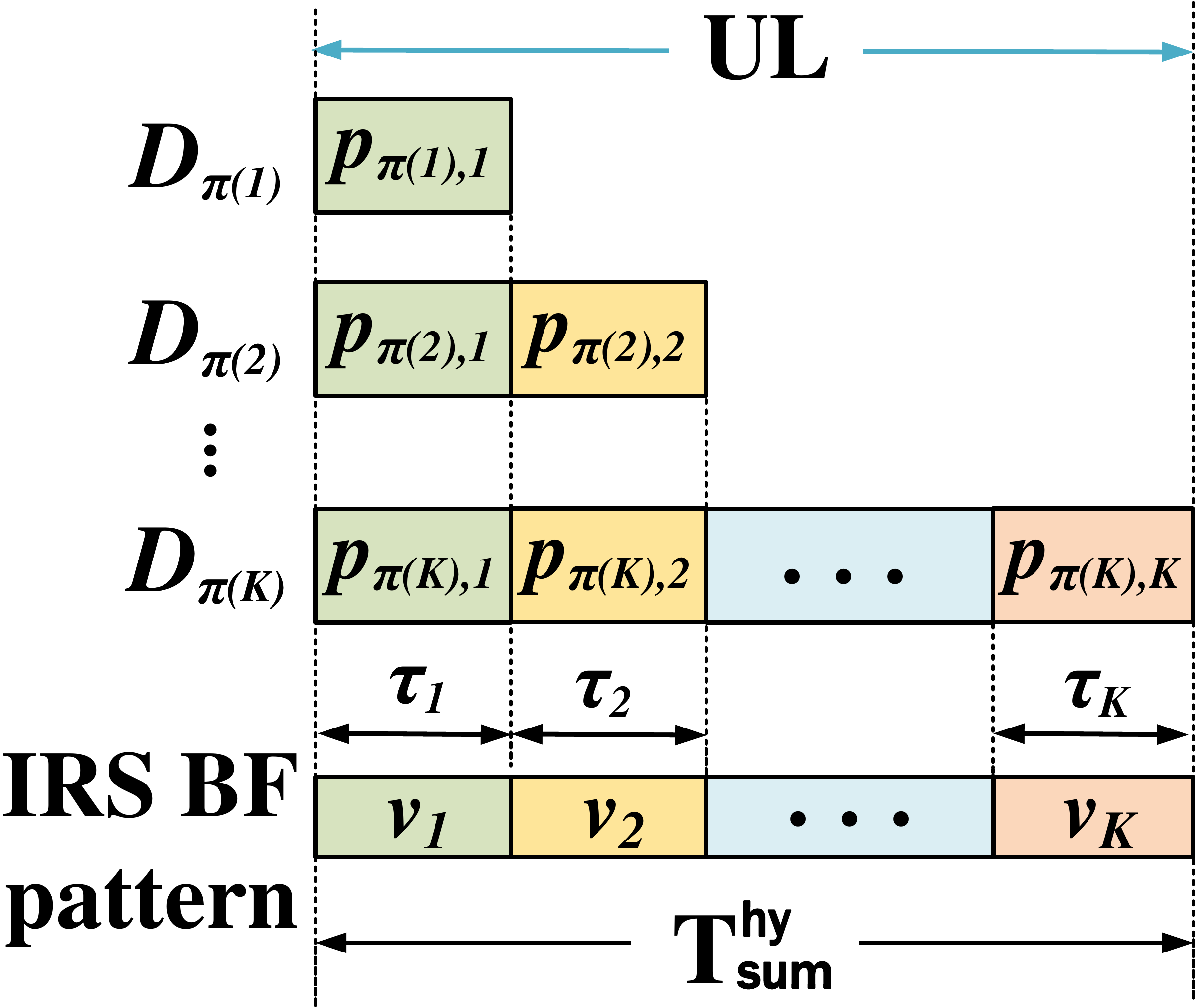}}\hspace{0.2cm}
    \subfigure[IRS-aided TDMA]{
        \label{fig:TDMA}
        \includegraphics[height=0.382\textwidth]{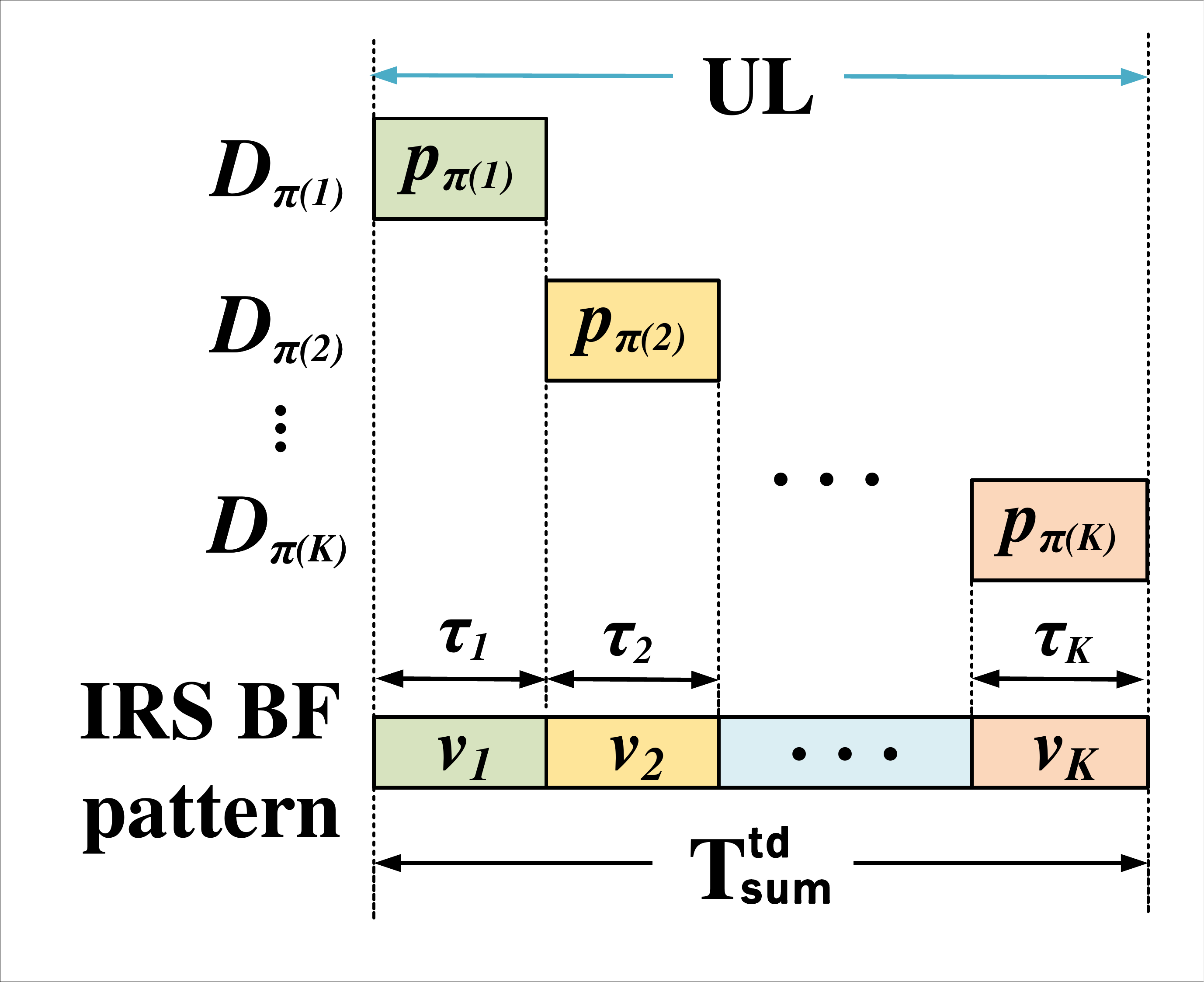}}
        \setlength{\abovecaptionskip} {-0.0cm}
        \caption{UL transmission protocols.} 
    \label{fig:protocol}
    \end{minipage}
    \end{figure}

    Motivated by the above issues, in this work, we focus on the MA design for an IRS-aided delay-sensitive UL transmission system where an IRS is deployed to assist the UL transmission from multiple IoT devices to a BS, as shown in Fig. \ref{fig:Channel}.
    To  fully reap the advantages of utilizing the IRS in minimizing the {transmission delay} of the entire UL transmission period, we propose {an} IRS-aided hybrid MA (HMA) UL transmission protocol by exploiting the flexibility in resource allocation and IRS BF of the TDMA protocol, as well as the high spectral efficiency of the NOMA protocol, where the device's rearranged order, time and power allocation, as well as IRS BF patterns are optimized jointly. 
    \textls[0]{It is worth noting that unlike the toy example of two-user based hybrid NOMA protocol shown in \cite{ding2018delay,zeng2019delay,zhou2020delay}, we present a more general multi-device HMA UL transmission protocol, where multiple devices can transmit their desired signals to the BS simultaneously. Furthermore, different from the design in \cite{ding2022hybrid,ding2018delay,zeng2019delay}, our proposed IRS-aided HMA (I-HMA) protocol takes the dynamic BF gain of IRS into account, which is able to reap additional advantages to minimize the sum transmission delay.
    Note that by properly designing the BF patterns over different time slots, the channels of all devices are no longer random and can be time-varying throughout each channel coherence block, 
    which is different from the conventional systems without IRS as investigated in \cite{ding2022hybrid,ding2018delay,zeng2019delay}. The time-division and resources reuse mechanism in our proposed I-HMA protocol provides more flexibility in the system design, which however imposes new challenges for the theoretical analysis and optimization. 
    The main contributions of this work are detailed as follows.}
    
    \begin{itemize}
    \item To exploit the time-selective channels introduced by dynamic IRS BF, as well as the high spectral-efficiency of NOMA protocol, {an} I-HMA protocol is developed for UL transmission by allowing multiple devices to transmit their desired signals to the BS simultaneously, which is different from the IRS-aided TDMA (I-TDMA) protocol. On the other hand, unlike the IRS-aided NOMA (I-NOMA) protocol, \textls[0]{which enforces all the devices to transmit concurrently during the entire period, the proposed protocol provides more flexibility for each device to choose its stop-time once it finishes its required transmission task and the possibility to vary its transmit powers as well as the IRS BF patterns among different time slots.} 
    \item Two typical communication scenarios are investigated in this work, namely the transmit power limited case and the transmit energy limited case\footnote{For power limited system, the devices are limited by a peak power constraint, such as conventional wireless communication scenarios. Whereas for energy limited systems, the devices only have a certain amount of energy, such as wireless sensor networks, or energy harvesting scenarios, or dedicated wireless power transmissions.}. 
    To shed light on the superiority of our proposed I-HMA protocol over the I-TDMA and I-NOMA protocols, we characterize the minimum achievable delay of the corresponding MA protocols and unveil the relationship among the proposed I-HMA protocol and the other two special cases (I-TDMA and I-NOMA), which provides a useful guideline to simplify the optimization.
    \item For the transmit power limited delay minimization, it is found that the maximum power transmission (MPT) strategy is the optimal solution for the last device under the proposed protocol. Besides, it is proved that the proposed I-HMA protocol always outperforms the I-TDMA protocol, whereas the proposed I-HMA protocol outperforms the I-NOMA protocol when the required transmission throughput for the latter devices is larger than a threshold. In contrast, for the transmit energy limited system, the proposed I-HMA protocol will reduce to the I-NOMA or I-TDMA protocol if the device's initial energy is sufficiently large or extremely low, respectively. We also derive the conditions under which the proposed I-HMA outperforms I-TDMA and I-NOMA.
    \item To solve the formulated challenging optimization problems under the proposed framework, we first provide a generic solution for the original problems, where an efficient device's ordering principle is proposed, and the other variables are optimized via an alternating optimization (AO)-based method. Specifically, with the given device's order, the original problem is decomposed into two problems, namely resource allocation optimization and IRS BF optimization. Based on the successive convex approximation (SCA) transformations and fractional programming (FP) method, each subproblem is optimized in an iterative way until convergence is achieved. Furthermore, we present the solution for the special cases, namely the I-TDMA and the I-NOMA based optimization problems.
    \item \textls[0]{Simulation results validate our theoretical findings and demonstrate the significant performance gains achieved by the proposed protocol. In particular, 1) it is shown that the performance gap between the proposed I-HMA and I-NOMA under the asymmetric deployment is wider than its counterpart under the symmetric deployment, which indicates that the NOMA protocol in our proposed I-HMA protocol plays a more prominent role under the asymmetric deployment; 2) it reveals that significant performance gain can be achieved by employing dynamic IRS BF compared with static IRS BF even with a small size of IRS, which indicates the effectiveness of the proposed dynamic IRS BF design; 3) it is observed that employing the proposed I-HMA protocol reduces the UL transmission time consumption not only from the system-centric aspect, but also from the device-centric aspect, which can reduce the energy consumption for each device.}
    \end{itemize}
    
    {\textls[1]{The rest of the paper is organized as follows. Section II introduces the system model and the problem formulations for both the transmit power limited system and the transmit energy limited system. Section III provides the performance analysis of the proposed I-HMA protocol compared with the  I-TDMA protocol and I-NOMA protocol. The proposed solution and algorithm  are provided in Section IV. Section V presents numerical results to evaluate the performance of the proposed design and provides some interesting insights. Finally, we conclude the paper in Section VI.}}

    \emph{Notations:} Throughout the paper, boldface lowercase letters and boldface uppercase letters represent vectors and matrices, respectively. Superscripts $(\cdot)^{{H}}$ represents the Hermitian transpose operator. $|\mbf{X}|$ denotes the absolute value of a scalar. $\mbb{C}^{a \times b}$ denotes the space of $a \times b$ complex matrices. $\mathscr{W}(\cdot)$ denotes the Lambert W function. $\Re\{\cdot\}$ represents the real part of a complex value. diag$(\cdot)$ denotes the diagonalization operator.
    
    \section{System Model And Problem Formulation}
    In this paper, we focus on an IRS-aided UL transmission system as shown in Fig. \ref{fig:Channel}, which consists of a single-antenna BS, an IRS with $N$ reflecting elements and $K$ single-antenna devices. 
    {Note that it can be readily extended to the case with multiple antennas at the BS by incorporating the receive beamforming design at the BS, which has been extensively studied in existing works.} The equivalent baseband channels from device $k$ to the IRS, from the IRS to the BS, and from device $k$ to the BS are denoted by $\bh_{r,k} \in \mbb{C}^{N \times 1}$, $\bg \in \mbb{C}^{N \times 1}$, and $h_{d,k} \in \mbb{C}$, $\forall k \in \mcal{K}\teq\{1,\cdots,K\}$, respectively. 
    {To characterize the maximum achievable performance, the channel state information (CSI) of the direct links and cascaded links is assumed to be available at the transceivers, which can be obtained by utilizing the techniques proposed in \cite{taha2021enabling,chen2019channel,wang2020channel,you2020channel,swi2022channel}. The extension to the case of the imperfect CSI will be left for future work.} Besides, all the wireless channels are assumed to be quasi-static flat-fading and remain constant within the transmission period. 
    
    \vspace{-0.2cm}
    \subsection{UL Transmission Protocol}\label{ sec:proto}
    \vspace{0.4cm}
     To exploit the favorable time-selective channels introduced by dynamic IRS BF as well as the high spectral efficiency of NOMA, in this work, we propose {an} I-HMA UL transmission protocol as shown in Fig. \ref{fig:HNOMA}, which combines the advantage of the  I-NOMA and I-TDMA protocols.\footnote{Existing techniques can be utilized for synchronization, whereas the investigation for the case of imperfect synchronization will be left as our future work.}
     This idea comes from the observation that the ``one device at a time'' mechanism in the  I-TDMA protocol will result in a relatively long delay for the entire UL transmission since the other devices need to wait until the current activated device finishes its transmission task.
     Specifically, under the  I-TDMA protocol as shown in Fig. \ref{fig:TDMA}, all the devices are arranged in a sequence to transmit their information data successively. To distinct such transmitting order from the device's original index, we use $\pi(\cdot)$ to denote the inverse function that maps the rearranged order index to the device's original index, and use $D_{\pi(k)}$ to represent the device  that in the $k$-th order to execute its transmission under the conventional TDMA protocol with the original index denoted by ${\pi(k)}$ as shown in Fig. \ref{fig:TDMA}. 
     
     To make the best use of the waiting time in TDMA-based UL transmission, we consider to introduce the NOMA technology for the UL transmission so as to enable simultaneous transmissions of multiple devices. 
     In particular, given the rearranged order ${\bm{\pi}} \teq \{ \pi(1), \cdots , \pi(K) \}$,  unlike the conventional TDMA protocol as shown in Fig. \ref{fig:TDMA}, where $D_{\pi(k)}$ needs to wait until the former $k-1$ devices, i.e., $D_{\pi(1)}, \cdots, D_{\pi(k-1)}$, finish their transmission and can only transmit during $\tau_k$, {the proposed HMA protocol allows $D_{\pi(k)}$ to utilize its waiting time, i.e., $\sum_{i=1}^{k-1} \tau_i$, as well as $\tau_k$ for UL transmission, which means that $D_{\pi(k)}$ can occupy at most a time duration of $\sum_{i=1}^k \tau_i$ for its transmission.
     And the user will  stop immediately after finishing its own transmission task.}
     Moreover, to reap the additional dynamic BF gain of the IRS as revealed in \cite{wu2021irs}, it is assumed that the IRS can reconfigure its reflecting elements dynamically among different time slots.\footnote{In this paper, we assume that the IoT devices are static or moving slowly, which is also one of the most typical scenarios for the application of IRS. In this case, the channel coherence time is about on the order of 25 milliseconds \cite{tse2005fundamentals}. In addition, as reported in \cite{li2022intelligent}, the minimum recovery time of PIN diode can be designed to be as small as a few nanoseconds, enabling a maximum radiation pattern changing speed with hundreds of MHz. 
     Therefore, it is practical to assume that the IRS reflection matrix can be reconfigured multiple times within the channel coherence time. }
     
     As such, compared to the  I-TDMA protocol, the proposed I-HMA protocol allows $K-(k-1)$ devices, i.e., $D_{\pi(k)}, \cdots, D_{\pi(K)}$, to transmit their desired information data to the BS simultaneously via the assistance of IRS in the $k$-th time slot. {On the other hand, compared to the  I-NOMA protocol, where all users start and end transmission simultaneously, the users may finish their transmission at different time in the proposed protocol. Therefore, the proposed protocol can provide more flexibility for each device to determine its transmission completion time and the possibility to vary its transmit powers as well as the IRS BF patterns among different time slots. Note that the transmission power of the devices remains the same during each time slot and only changes among different time slots. And the BS will decode each device’s signal successively by utilizing SIC in each time slot.} 
     \subsection{Achievable Throughput}
     For our proposed I-HMA UL transmission protocol, the signal received at the BS with the aid of the IRS in the $i$-th time slot can be expressed as
     \begin{align}
        y_i &= \sum_{k=i}^K \sqrt{p_{\pi(k),i}} ( h_{d,\pi(k)} + \bg^{H} \IRS_i \bh_{r,\pi(k)} ) s_{\pi(k),i}+ n_{i} =   \sum_{k=i}^K \sqrt{p_{\pi(k),i}}  \bb_{\pi(k)}^{H} \tv_i   s_{\pi(k),i}+ n_{i},
     \end{align}\noindent where  $\bb_{\pi(k)}\in \mbb{C}^{(N+1) \times 1}$ denotes the composite channel of $D_{\pi(k)}$, which is given by $\bb_{\pi(k)}^H \teq [\bg^H \dia(\bh_{r,\pi(k)}),h_{d,\pi(k)}]$. $\IRS_i\teq \dia([e^{j\theta_{i,1}},\cdots,e^{j\theta_{i,N}}])$ denotes the diagonal IRS BF coefficient matrix in the $i$-th time slot. And $\tv_i$ is defined as $\tv_i^H\teq[e^{-j\theta_{i,1}},\cdots,e^{-j\theta_{i,N}},1] $. Besides, $p_{\pi(k),i}$ and $s_{\pi(k),i}$ denote the transmit power and signal of $D_{\pi(k)}$ in the $i$-th time slot, respectively. $n_{i}$ denotes the additive white Gaussian noise with the power denoted by $\noi$.
     
     \textls[1]{To decode ${\pi(k)}$-th device's signal, $\forall k \in \{i,\cdots,K\}$, from the composite signals from $D_{\pi(i)}, \cdots,$ $D_{\pi(K)}$  in the  $i$-th time slot, the SIC is employed at the BS\footnote{Regarding the case of imperfect SIC, the proposed algorithm is still applicable by making some transformation. We will make a deeper investigation to this scenario in our future work.}. And it is assumed that the BS first decodes the signal of the device with a larger order so as to ensure the device with prior order experiences smaller interference, i.e., the BS decodes $D_{\pi(j)}$'s signal before decodes $D_{\pi(i)}$'s signal, $\forall j>i$. As such, the achievable throughput of $D_{\pi(k)}$ in bits in the  $i$-th time slot $\tau_i$ can be calculated as}
     \begin{align}
     \tau_i \hspace{0.05cm} r_{\pi(k),i}=
        \begin{cases}\tau_i B \log_2 \left( 1+ \frac{ p_{\pi(k),i} | \bb_{\pi(k)}^{H} \tv_i |^2 }{\sum_{j=i}^{k-1}  p_{\pi(j),i} | \bb_{\pi(j)}^{H} \tv_i |^2 + \noi} \right),   &i<k,  \\
        \tau_i B \log_2 \left( 1+ { p_{\pi(k),i} | \bb_{\pi(k)}^{H} \tv_i |^2 }/{\noi} \right), & i=k,
        \end{cases} \label{eq:rate}
    \end{align}\noindent where $B$ denotes the transmission bandwidth. It can be seen in (\ref{eq:rate}) that, since the BS decodes all the other device's signals before decoding the ${\pi(k)}$-th device's signal for the signals received in the $k$-th time slot, $D_{\pi(k)}$ will experience interference-free transmission as that in TDMA during $\tau_k$. On the other hand, it is worth noting that $D_{\pi(k)}$ can occupy at most a time duration of $\sum_{i=1}^k \tau_i$ for transmission with the proposed I-HMA protocol. Thus, its total achievable throughput can be calculated as $\sum_{i=1}^k \tau_i r_{{\pi(k)},i}$.
    
    \subsection{Problem Formulations}\label{sec:Problem_formu}
    \vspace{0.3cm}
    Without loss of generality, two typical communication scenarios are investigated, namely the \tbf{1) transmit power limited case} and \tbf{2) the transmit energy limited case}. 
    For brevity, denote $ \mcal{N} \teq  \{ 1,\cdots ,N\} $, $ \mcal{I}(k) \teq  \{ 1,\cdots ,k\} $, and ${\bm{\Pi}}$ as the set of all possible device's rearranged order with $|{\bm{\Pi}}|=K!$. Besides, let ${\bm{p}} \teq \{ p_{\pi(k),i} \hspace{0.05cm} | \hspace{0.1cm} \forall i \in \mcal{I}(k) , \forall k \in \mcal{K}\}$, ${\bm{\Psi}} \teq \{ \tv_1,\cdots,\tv_{K}\}$, and ${\bm{\tau}} \teq \{ \tau_1, \cdots , \tau_{K} \}$. 
    Our goal is to minimize the sum transmission delay during the whole UL transmission frame, by jointly optimizing the device's rearranged order ${\bm{\pi}}$, time allocation ${\bm{\tau}}$, transmit power ${\bm{p}}$ and IRS BF patterns set ${\bm{\Psi}}$. 
    
    \tbf{1) Transmit Power Limited Delay Minimization:}
    For the transmit power limited case, each device is powered by a stable source, and its transmit power is limited by the source's maximum output power $P_{\pi(k)}^{\m}$ for $D_{\pi(k)}$. The corresponding optimization problem is formulated as
    \begin{subequations}
    \begin{align}
    (\text{P} 1): \hspace{0.2cm} \underset{{\bm{\pi}}, {\bm{\tau}}, {\bm{p}} ,{\bm{\Psi}} }{\min} \hspace{0.2cm} &  \sum\nolimits_{i=1}^K \tau_{i}  \label{eq:obj_P1}\\
    \text { s.t. } \hspace{0.2cm}   &\sum\nolimits_{i=1}^k \tau_i r_{{\pi(k)},i} \geq L_{\pi(k)}, \forall k \in \mcal{K},\label{eq:con_Lk_P1} \\
    & 0\leq p_{{\pi(k)},i} \leq P_{\pi(k)}^{\m},  \forall i \in \mcal{I}(k) , \forall k \in \mcal{K},   \label{eq:con_p_P1} \\
    & \tau_{i} \geq 0, \forall i \in \mcal{K}, \label{eq:con_t_noneg_P1}\\
    & \left|\left[ \tv_i \right]_{n}\right|=1,\forall n \in \mcal{N}, \forall i \in \mcal{K}, \label{eq:con_v_P1}\\
    & {\bm{\pi}} \in {\bm{\Pi}}\label{eq:con_order_P1}.
    \end{align}
    \end{subequations}
    
    \vspace{-0.4cm}
    
    \noindent \textls[1]{In (P1), (\ref{eq:con_Lk_P1}) denotes the quality-of-service (QoS) constraint, where $L_{\pi(k)}$ denotes the ${\pi(k)}$-th device's throughput requirement for the UL transmission. 
    (\ref{eq:con_p_P1}) and (\ref{eq:con_t_noneg_P1}) are the transmit power constraints and the non-negativity constraints on the optimization variables, respectively.} (\ref{eq:con_v_P1}) denotes the unit-modulus constraints of the IRS phase-shifts vectors. (\ref{eq:con_order_P1}) is the device's order constraint.

    \tbf{2) Transmit Energy Limited Delay Minimization:}
    For the transmit energy limited case, a certain amount of energy $E_{\pi(k)}$ Joule (J) for device $D_{\pi(k)}$ is available at the beginning of each transmission period which can be used for its UL transmission. The corresponding optimization problem can be similarly formulated as
    \begin{subequations}
    \begin{align}
    (\text{P} 2): \hspace{0.2cm}\underset{ {\bm{\pi}},{\bm{\tau}}, {\bm{p}} ,{\bm{\Psi}}  }{\min} \hspace{0.2cm} &  \sum\nolimits_{i=1}^K \tau_{i}  \label{eq:P2_obj}\\
    \text { s.t. } \hspace{0.2cm}   &(\ref{eq:con_Lk_P1}),(\ref{eq:con_t_noneg_P1}),(\ref{eq:con_v_P1}),(\ref{eq:con_order_P1})\label{eq:P2_con1} \\
    &\sum\nolimits_{i=1}^k p_{{\pi(k)},i} \tau_i \leq  E_{\pi(k)}, \forall k \in \mcal{K}, \label{eq:P2_con2} \\
    &  p_{{\pi(k)},i} \geq 0,  \forall i \in \mcal{I}(k) , \forall k \in \mcal{K}   \label{eq:P2_con3} , 
    \end{align}
    \end{subequations}
    
    \vspace{-0.3cm}
    
    \noindent where (\ref{eq:P2_con2}) and (\ref{eq:P2_con3}) are the transmit energy constraint and the non-negativity constraints on the optimization variables, respectively.
    
    It is worth noting that for the system with $K$ devices, there are $K!$ rearranged orders in total. Moreover, with the assistance of the IRS, the equivalent channels vary with different IRS BF patterns, which
    makes the design of the device's rearranged order more difficult \cite{mu2020exploiting,mu2021joint}. Generally, there are no efficient solutions to determine the optimal device's order even in the conventional communication system without IRS. As such, obtaining the optimal device's rearranged order requires an exhaustive search over every possible user ordering, which needs to execute $K!$ times optimization and is computationally prohibitive with the increase of $K$. {Apart from the difficulty in determining the optimal device's rearranged order, (P1) and (P2) are non-convex due to highly coupled optimization variables involved in the constraint (\ref{eq:con_Lk_P1}). Moreover, the variables set ${\bm{\tau}}$ in the objective function is actually an implicit function with respect to (w.r.t.) ${\bm{p}}$ and ${{\bm{\Psi}}}$. As such, problem (P1) and (P2) are difficult to solve optimally in general. Nevertheless, in the following, we will present an efficient algorithm to obtain their high-quality solutions.} For simplicity, we denote $\bar{L}_{\pi(k)}$ as the normalized required transmission throughput in bits/Hz, which is given by $\bar{L}_{\pi(k)}=L_{\pi(k)}/B$, $\forall k \in \mcal{K}$, and denote $\gamma_k(\tv)\teq | \bb_{{\pi(k)}}^{H} \tv |^2/\noi$.
    
    \section{Performance Analysis}
    It is worth noting that both  I-TDMA and  I-NOMA can be viewed as special cases of the proposed I-HMA protocol. Specifically, on one hand, by setting $p_{{\pi(j)},i}=0$, $\forall i \in \mcal{I}(k-1) $, $\forall k \in \mathcal{K}$, the proposed I-HMA protocol reduces to the  I-TDMA protocol. Whereas, by setting $\tau_i=0$, $\forall i \in\{2,\cdots,K\}$, the proposed I-HMA protocol degrades to the  I-NOMA protocol. On the other hand, the optimal solution to the problem with  I-TDMA and I-NOMA is always feasible for the problem with the proposed I-HMA protocol. Thus, the performance of the proposed I-HMA protocol is no worse than the other two benchmarks. However, it remains an open problem whether the proposed I-HMA protocol is actually beneficial for minimizing the sum transmission delay of the whole UL transmission compared to the other two  strategies (I-TDMA and I-NOMA). To answer this question, in this section, we analyze the performance of the proposed I-HMA protocol compared with the  I-TDMA and  I-NOMA protocol under the transmit power limited case and transmit energy limited case, respectively.
    
    In the following,  it is assumed that the analysis is under the given order ${\bm{\pi}}$. Besides, we suppose that $\mathscr{S}^{\hy ^*}({\bm{\Psi}})\teq \{{\bm{\tau}}^{\hy ^*}, {\bm{p}}^{\hy ^*} |\hspace{0.05cm} {\bm{\Psi}}  \}$, $\mathscr{S}^{\td ^*}({\bm{\Psi}})\teq \{{\bm{\tau}}^{\td ^*}, {\bm{p}}^{\td ^*}|\hspace{0.05cm} {\bm{\Psi}} \}$, and $\mathscr{S}^{\no ^*}( {\bm{\Psi}})\teq \{{\bm{\tau}}^{\no ^*}, {\bm{p}}^{\no ^*} |\hspace{0.05cm} {\bm{\Psi}}  \}$\footnote{For I-NOMA based solution, it follows that ${\bm{\tau}}^{\no }=\tau_1$ and ${\bm{\Psi}}^{\no }=\tv_1$.} achieve the optimal solution to the problem with the proposed I-HMA protocol, with the  I-TDMA protocol, and with the  I-NOMA protocol with the IRS BF patterns set ${\bm{\Psi}}$, respectively. 
    $\mathsf{T_{sum}^{hy}}({\bm{\Psi}})$, $ \mathsf{T_{sum}^{td}}({\bm{\Psi}})$, and $ \mathsf{T_{sum}^{no}}({\bm{\Psi}})$ denote the sum transmission delay achieved at $\mathscr{S}^{\hy ^*}({\bm{\Psi}})$, $\mathscr{S}^{\td ^*}({\bm{\Psi}})$, and $\mathscr{S}^{\no ^*}({\bm{\Psi}})$, respectively. $\mathsf{T_{sum}^{hy^*}}$, $\mathsf{T_{sum}^{td^*}}$ and $\mathsf{T_{sum}^{no^*}}$ denote the minimum sum transmission delay of employing the proposed I-HMA protocol, I-TDMA protocol and I-NOMA protocol for UL transmission, respectively.

    \subsection{Transmit Power Limited Case}
    The optimal performance between the proposed I-HMA protocol and the  I-TDMA protocol can be compared as follows.

    \begin{Pro}\label{pro:HMA_sup}
        For the transmit power limited system, we have
    \begin{subequations}
    \begin{numcases}{}
     \mathsf{T_{sum}^{hy}}({\bm{\Psi}})< \mathsf{T_{sum}^{td}}({\bm{\Psi}}),\\
     \mathsf{T_{sum}^{hy^*}}=\mathsf{T_{sum}^{hy}}({\bm{\Psi}}^{\hy^*})<\mathsf{T_{sum}^{hy}}({\bm{\Psi}}^{\td^*})<\mathsf{T_{sum}^{td}}({\bm{\Psi}}^{\td^*}) =\mathsf{T_{sum}^{td^*}}\label{eq:IRD_DBF_gain}.
    \end{numcases}
    \end{subequations}
    \end{Pro}
    
    \begin{myproof}
    Please refer to Appendix A.
    \end{myproof}
    
    {Proposition \ref{pro:HMA_sup} shows that given the same ${\bm{\Psi}}$, the sum transmission delay of employing the I-TDMA protocol is always larger than that with the proposed I-HMA protocol for UL transmission under the transmit power constraints, which indicates that the proposed I-HMA protocol always outperforms the I-TDMA protocol in this scenario. Moreover, with the assistance of the IRS, the performance gap between the proposed I-HMA protocol and the  I-TDMA protocol becomes larger as indicated by (\ref{eq:IRD_DBF_gain}), since introducing the IRS provides additional DoFs and brings additional gain by properly optimizing the IRS BF patterns.}
	
	For the performance comparison between the proposed I-HMA protocol and the  I-NOMA protocol, we have the following proposition.
	\begin{Pro}\label{pro:HMA_eq_NOMA}
        For the transmit power limited system, if ${L}_{\pi(k)} \leq   {L}_{\pi(k)}^{{\no}}( \tv_1) ,\forall k>1$, then it follows that
        \begin{equation}
        \mathsf{T_{sum}^{hy}}( {\bm{\Psi}}) = \mathsf{T_{sum}^{no}}( {\bm{\Psi}})={\bar{L}_{\pi(1)}}/\hspace{0.03cm}{\log_2 \left[1+P_{\pi(1)}^{\m}\gamma_1(\tv_1)\right]},
        \end{equation}\noindent  where ${L}_{\pi(k)}^{{\no}}(\tv_1)$ is defined as
        \begin{equation}
        \begin{aligned}
       {L}_{\pi(k)}^{\no}(\tv_1)\teq  {\log_2 \left[1+\frac{P_{{\pi(k)}}^{\m}{\gamma}_k(\tv_1)}{1+\sum_{i=1}^{k-1} P_{{\pi(i)}}^{\m}{\gamma}_i(\tv_1)} \right]}{L}_{\pi(1)}\bigg/\hspace{0.03cm}{\log_2 \left[1+P_{\pi(1)}^{\m}\gamma_1(\tv_1)\right]} .
        \end{aligned}\label{eq:L_bound}
        \end{equation}
    \end{Pro}
    
    \begin{myproof}
    Please refer to Appendix B.
    \end{myproof}
    
    Proposition \ref{pro:HMA_eq_NOMA} indicates that compared with $L_{\pi(1)}$, if the required transmission throughput for the latter devices, i.e., $L_{\pi(k)}$, $\forall k>1$, are relatively small such that can be completed within the first time slot even when all the devices transmit at their maximum power, the proposed I-HMA protocol reduces to  I-NOMA protocol, and results in the same optimal solution. In this case, the minimum sum transmission delay is limited by the $D_{\pi(1)}$. 
    Besides, it can be seen from (\ref{eq:L_bound}) that the BF patterns of IRS, i.e., $\tv_1$, does influence the threshold for the performance gap between the proposed I-HMA protocol and I-NOMA protocol, which validates the  effect of IRS.
    Proposition \ref{pro:HMA_eq_NOMA} provides a region where the performance of the proposed I-HMA protocol is the same as the  I-NOMA protocol. 
    However, if the condition in Proposition \ref{pro:HMA_eq_NOMA} does not hold, whether the proposed I-HMA protocol outperforms the  I-NOMA protocol or not remains an open question since the former devices may either reduce their transmit power so as to guarantee all the devices finish their transmission simultaneously, which is the  I-NOMA protocol; or choose the proposed I-HMA protocol. To answer this question, next, we show the superiority of the proposed I-HMA protocol when the condition in Proposition \ref{pro:HMA_eq_NOMA} does not hold by taking the two-device system as an example.
    \begin{Pro}\label{pro:HMA_sup_no}
        For the transmit power limited system with two devices, if ${L}_{\pi(2)} >  {L}_{\pi(2)}^{{\no}}( \tv_1)$, where ${L}_{\pi(2)}^{{\no}}( \tv_1)$ is given by (\ref{eq:L_bound}), then we have
    \begin{subequations}
    \begin{numcases}{}
     \mathsf{T_{sum}^{hy}}( {\bm{\Psi}}) < \mathsf{T_{sum}^{no}}( {\bm{\Psi}}),\\
     \mathsf{T_{sum}^{hy^*}}=\mathsf{T_{sum}^{hy}}({\bm{\Psi}}^{\hy^*})<\mathsf{T_{sum}^{hy}}({\bm{\Psi}}^{\no^*})<\mathsf{T_{sum}^{no}}({\bm{\Psi}}^{\no^*}) =\mathsf{T_{sum}^{no^*}}.\label{eq:NOMA_po_IRS}
    \end{numcases}
    \end{subequations}
    \end{Pro}
    
    \begin{myproof}
    Please refer to Appendix C.
    \end{myproof}
    
    Proposition \ref{pro:HMA_sup_no} shows that given the same ${\bm{\Psi}}$, the sum transmission delay of employing the I-NOMA protocol is larger than that with the proposed I-HMA protocol for UL transmission under the maximum power constraints if ${L}_{\pi(k)} \leq   {L}_{\pi(k)}^{{\no}}( \tv_1) ,\forall k>1$. Moreover, after optimizing the IRS BF patterns properly, the performance gain brought by the proposed I-HMA protocol will further enlarge, as indicated by (\ref{eq:NOMA_po_IRS}), which shows the effect of introducing the IRS.

    Besides, Propositions \ref{pro:HMA_eq_NOMA} and \ref{pro:HMA_sup_no} reveal that under the transmit power constraints, the optimal performance of the proposed I-HMA protocol has a threshold-based relationship compared with the  I-NOMA protocol under the same device's order, which can simplify the optimization if the condition in Proposition \ref{pro:HMA_eq_NOMA} holds. Specifically, we can first solve the NOMA-based optimization by setting $\tau_j=0$, and $p_{{\pi(j)},i}=0$, $\forall i \in \{2, \cdots,j\}$, $\forall j \in\{2,\cdots,K\}$, in (P1). As such, only $\tau_1$, $\tv_1$, and $p_{{\pi(k)},1}$, $\forall k \in\mcal{K}$ need to be optimized for this special case. After obtaining $\mathscr{S}^{\no ^*}({\bm{\Psi}}^{\no ^*})$, we can examine whether the condition in Proposition \ref{pro:HMA_eq_NOMA} is satisfied or not. If so, we can directly obtain the optimal solution of $\mathscr{S}^{\hy ^*}({\bm{\Psi}}^{\hy ^*})$ with the aforementioned $\mathscr{S}^{\no ^*}({\bm{\Psi}}^{\no ^*})$ instead of solving the complex multi-variable problem (P1). Thus, the computational complexity can be significantly reduced.

    \subsection{Transmit Energy Limited Case}
    
    \begin{Pro}
    \label{pro:bound}
    For the transmit energy limited case, we have
    \begin{enumerate}
    \item if ${E}_{\pi(k)} \geq  {E}_{\pi(k)}^{{\no}}(\tv_1)$, $\forall k >1$, then $\mathsf{T_{sum}^{hy}}({\bm{\Psi}}) = \mathsf{T_{sum}^{no}}({\bm{\Psi}})=\tau_1^{\no ^*}=\mathscr{G}_1(\tv_1)$;
    \item if ${E}_{\pi(k)}^{{\no}}(\tv_1)>  {E}_{\pi(k)}>  {E}_{\pi(k)}^{{\td}}({\bm{\Psi}})$, $\forall k >1$, then $\mathsf{T_{sum}^{hy}}({\bm{\Psi}}) < \min\{ \mathsf{T_{sum}^{td}}({\bm{\Psi}}), \mathsf{T_{sum}^{no}}({\bm{\Psi}})\}$;
    \item if ${E}_{\pi(k)}^{{\td}}({\bm{\Psi}})\geq  {E}_{\pi(k)} \geq 0$, $\forall k >1$, then $\mathsf{T_{sum}^{hy}} = \mathsf{T_{sum}^{td}}({\bm{\Psi}})=\sum_{k=1}^K \tau_{k}^{\td ^*}=\sum_{k=1}^K \mathscr{G}_k(\tv_k)$.
    \end{enumerate}

    \noindent Specifically, the threshold ${E}_{\pi(k)}^{{\no}}(\tv_1)$ is computed as
        \begin{equation}
        \begin{aligned}
         {E}_{\pi(k)}^{{\no}}(\tv_1) & \teq \frac{\mathscr{G}_1(\tv_1) }{{\gamma}_k(\tv_1)} \left( 2^{{\bar{L}_{\pi(k)}}/{\mathscr{G}_1(\tv_1)}  } -1\right) 2^{\sum_{j=1}^{k-1} {\bar{L}_{\pi(j)}}/{\mathscr{G}_1(\tv_1)}}.
        \end{aligned}\label{eq:E_no}
        \end{equation}\noindent Whereas, the threshold ${E}_{\pi(k)}^{{\td}}({\bm{\Psi}})$ is calculated as %
   \begin{equation}
        \begin{aligned}
       \hspace{-0.2cm} E_{\pi(k)}^{\td }({\bm{\Psi}}) & \teq \frac{\bar{L}_{\pi(k)}\left[ {2^{{\bar{L}_{\pi(i)}} /{\mathscr{G}_i(\tv_i)}}}\big/{\gamma_k(\tv_i)} - {1}\big/{\gamma_k(\tv_k)} \right]}{\log_2 \left[ {\gamma_k(\tv_k)}/{\gamma_k(\tv_i)}  \right] + {\bar{L}_{\pi(i)}}/{\mathscr{G}_i(\tv_i)}},
        \end{aligned}\label{eq:E_td}
        \end{equation}
        
        \noindent with $i$ and $\mathscr{G}_k(\tv_k)$, $\forall k \in \mcal{K}$, respectively given by
         \begin{equation}
        \begin{aligned}
        i & =  \underset{  j \in \mcal{I}(k-1)   }{\arg \min}  {\left[1+p_{{\pi(j)},j}^{\td ^*}(\tv_j )\gamma_j(\tv_j )\right] }\big/{\gamma_k(\tv_j )},
        \end{aligned}\label{eq:E_td_i}
        \end{equation}\noindent 
        \begin{equation}
        \begin{aligned}
       \mathscr{G}_k(\tv_k)= -\frac{E_{\pi(k)}{\gamma}_k(\tv_k)\bar{L}_{\pi(k)}\ln 2}{E_{\pi(1)}{\gamma}_k(\tv_k)\mathscr{W}\left(\xi_k(\tv_k) e^{\xi_k(\tv_k)} \right)+\bar{L}_{\pi(k)}\ln 2},
        \end{aligned}\label{eq:TDMA_E}
        \end{equation}\noindent where $\mathscr{W}(\cdot)$ denotes the Lambert W function \cite{corless1996lambertw}, and
    \begin{equation}
        \begin{aligned}
       \xi_k(\tv) &= -{\bar{L}_{\pi(k)}\ln 2}\big/\left[{E_{\pi(k)}{\gamma}_k(\tv)}\right].
        \end{aligned}\label{eq:xi_td}
        \end{equation}
    \end{Pro}
    
    \begin{myproof}
    Please refer to Appendix D.
    \end{myproof}
    
    Proposition \ref{pro:bound} indicates that the proposed I-HMA protocol outperforms the I-TDMA and the I-NOMA protocol when device's initial energy is in a region. Whereas the proposed I-HMA protocol will reduce to the other two protocols when  device's initial energy is sufficiently large or not sufficient, which can simplify the optimization. 
    Specifically, we can first obtain the optimal solution of the problem for the two special cases, i.e., obtain $\mathscr{S}^{\td ^*}({\bm{\Psi}}^{\td *})$ by solving the I-TDMA based optimization; and obtain the optimal $\tv_1^{\no^*}$ by solving the I-NOMA based optimization. Afterwards, we can calculate $ E_{\pi(k)}^{\no }(\tv_1^{\no^*})$ and $ E_{\pi(k)}^{\td }({\bm{\Psi}}^{\td *})$ with the obtained $\tv_1^{\no^*}$ and $\mathscr{S}^{\td ^*}({\bm{\Psi}}^{\td *})$. If the condition of case 1 or case 3 in Proposition \ref{pro:bound} holds, then we can directly obtain the optimal solution of $\mathscr{S}^{\hy ^*}({\bm{\Psi}}^{\hy ^*})$ with the aforementioned $\mathscr{S}^{\no ^*}({\bm{\Psi}}^{\no ^*})$ or $\mathscr{S}^{\td ^*}({\bm{\Psi}}^{\td ^*})$ instead of solving the original problem (P2), which can reduce the number of variables to be optimized.
   \begin{Rem}
       It is worth noting that the case in \cite{ding2018delay} and \cite{zeng2019delay} can be viewed as a special case of our considered system by setting $K=2$, and leaving out the influence of the IRS.
       Moreover, the results in \cite{ding2018delay} and \cite{zeng2019delay} cannot be extended to our considered system, due to the following two reasons. First, simultaneously serving multiple users imposes additional DoFs and also introduces new challenges for deriving the thresholds as presented in Proposition \ref{pro:bound}, especially for the derivation of the threshold ${E}_{\pi(k)}^{{\td}}$, where the expressions obtained by utilizing the KKT conditions as given in Appendix D are much more complex than that of the two-user case. Second, with the assistance of the IRS, especially under the dynamic IRS BF configuration, the channels for each device vary among different time slots, which further influences the aforementioned thresholds and the existing conclusions cannot be applied directly.
   \end{Rem}

   \section{Proposed Algorithms for (P1) and (P2)}\label{sec:solu}
   As discussed in Section \ref{sec:Problem_formu}, (P1) and (P2)\footnote{Note that (P1) is always feasible since the transmission time can be prolonged to guarantee the QoS constraints (\ref{eq:con_Lk_P1}). Whereas, (P2) is feasible if $E_{\pi(k)}^{\min \star} \leq E_{\pi(k)}$, $\forall k\in\mcal{K}$, where $E_{\pi(k)}^{\min \star}$ is the minimum required energy of $D_{\pi(k)}$, which can be obtained by solving the energy minimization problem without transmission delay constraint.} are complex non-convex optimization problems with closely coupled variables, which generally cannot be solved optimally. 
   Nevertheless, after deeply exploiting the special structure of the optimal solution for (P1) and (P2), respectively in Section III, we can simplify the optimization by examining the conditions as indicated by Propositions \ref{pro:HMA_eq_NOMA} and \ref{pro:bound}, and then resorting to only the I-TDMA based optimization or I-NOMA based optimization problems, which are much simpler to solve. 
   Therefore, in the following, we first propose a generic algorithm for (P1) and (P2), respectively. Then, we present the algorithm for the special cases of the I-TDMA based and I-NOMA based optimization problems, respectively.
    
   \subsection{Proposed Algorithm for (P1)}\label{sec:solu_HMA_po}
   \vspace{0.3cm}
    To cope with the multi-variable coupled challenging problem, we first propose a principle to rearrange the device's order. Then, we solve the problem with the optimized order by decomposing the original problem (P1) into two subproblems, namely the resource allocation optimization problem and the IRS BF optimization problem, which are solved respectively in an alternative manner until the objective value converges.
   
   \tbf{1) Device's Ordering Principle:}
   To begin with, we first deal with the device's ordering problem.
   It is worth noting that, different from the special case with the TDMA protocol where the device's rearranged order will not affect the sum transmission delay of the system, this order will influence the sum transmission delay with the proposed I-HMA UL transmission protocol due to the integration of the NOMA protocol.
   For the system with $K$ devices, there are $K!$ rearranged orders, which is prohibitive for exhaustive search when $K$ is large. Since there are no explicit expressions to evaluate the effect of device's rearranged order on the minimum sum transmission delay, it is hard to determine the optimal device's order even in the conventional communication system without IRS. 
   Nevertheless, in the following, we provide an effective principle to determine the device's order.

   To draw useful insights into the device's ordering of the multiuser case, we first investigate a special case of the system with two devices. In this case, there are only two possible rearranged orders. We use the superscript $\land$ to represent $D_1$ is ranked before $D_2$, whereas utilizing the superscript $\lor$ to represent $D_2$ is ranked before $D_1$, i.e., $\hat{\pi}(1)=\check{\pi}(2)=1$, $\hat{\pi}(2)=\check{\pi}(1)=2$.  Thus, $\hat{\gamma}_1(\cdot)=\check{\gamma}_2(\cdot)\teq \gamma_1(\cdot)$, $\hat{\gamma}_2(\cdot)=\check{\gamma}_1(\cdot)\teq \gamma_2(\cdot)$. For simplicity, we assume that $\tv_k$, $\forall k \in \mcal{K}$ is given by aligning $D_{\pi(k)}$'s IRS-reflected signal with non-IRS-reflected signal, i.e., $ \big[\tv_k^{\td ^*}\big]_n=e^{j\angle{\left[\bb_{\pi(k)}\right]_n}},\forall n \in \mcal{N}$. Therefore, we have $\hat{\tv}_1=\check{\tv}_2\teq\tv_1$ and $\hat{\tv}_2=\check{\tv}_1\teq\tv_2$. {As shown in Appendix C, the optimal power allocation strategy yields $\hat{p}_{1,1} = \check{p}_{1,1} = \check{p}_{1,2} = P_1^{\m}$ and $\hat{p}_{2,1} = \hat{p}_{2,2} = \check{p}_{2,1} = P_2^{\m}$.} Therefore, the minimum sum transmission delay of these two decoding orders can be compared as 
    \begin{equation}
		\begin{aligned}
		 &\hspace{-0.3cm}\mathsf{\hat{T}_{sum}^{hy^*}}-\mathsf{\check{T}_{sum}^{hy^*}}= \frac{\bar{L}_2 \log_2 \left[1+\frac{P_1^{\m}{\gamma}_1(\tv_2)}{1+P_2^{\m}{\gamma}_2(\tv_2)} \right] - \bar{L}_1 \log_2 \left[1+\frac{P_2^{\m}{\gamma}_2(\tv_1)}{1+P_1^{\m}{\gamma}_1(\tv_1)} \right]}{\log_2 [1+ P_1^{\m}{\gamma}_1(\tv_1)] \log_2 [1+ P_2^{\m}{\gamma}_2(\tv_2)]}.
		\end{aligned}\label{eq:compare}
    \end{equation}
    
    \noindent It can be seen that $D_1$ should be arranged before $D_2$ when $\bar{L}_1 / \log_2 \left[1+\frac{P_1^{\m}{\gamma}_1(\tv_2)}{1+P_2^{\m}{\gamma}_2(\tv_2)} \right] > \bar{L}_2 /$ $ \log_2 \left[1+\frac{P_2^{\m}{\gamma}_2(\tv_1)}{1+P_1^{\m}{\gamma}_1(\tv_1)} \right] $, whereas $D_2$ should be arranged first otherwise.

    Inspired by the analysis on the optimal rearranged order of the two-device case, we propose an efficient device ordering principle as follows. Specifically, we first calculate the optimal $ {\bm{p}}^{\td ^*}$, ${\bm{\tau}}^{\td ^*}$ and ${\bm{\Psi}}^{\td ^*}$. Then, we can compute the TDMA-based signal-to-noise ratio (SNR) denoted as $\varrho^{\td}_{\pi(k)},\forall k \in \mcal{K}$, by
    {\setlength\abovedisplayskip{-0.1cm}
    \begin{align}
		 \varrho^{\td}_{\pi(k)}=p_{\pi(k)}^{\td^*}{\gamma}_k\left(\tv^{\td^*}_k \right), \hspace{0.2cm}\forall k \in \mcal{K}.\label{eq:SNR_P}
    \end{align}}\noindent Afterwards, we can rearrange the devices according to the ascending order of the TDMA-based SNR ${\bm\varrho}^{\td}\teq\{\varrho^{\td}_{1},\cdots, \varrho^{\td}_{K}\}$, i.e., update ${\bm{\pi}}$ to guarantee $\varrho_{\pi(1)}^{\td} \leq  \cdots \leq \varrho_{\pi(K)}^{\td}$.

    The above device ordering principle is effective for the general case with multiple devices in the sense that it increases both of the value and weight for the average sum rate of each time slot.
    To illustrate this, we introduce the concept of sum rate during each time slot denoted by $R_{\su}^{(i)}$, $\forall i \in \mcal{K}$, and the concept of average sum rate during the entire UL transmission period denoted by $ \overline{R}_{\su}$, which are respectively expressed as
    \begin{align}
    R_{\su}^{(i)}=&\hspace{0.1cm}\sum_{k=i}^K r_{{\pi(k)},i} = B \log_2 \bigg( 1+ \sum_{k=i}^K \varrho_{\pi(k),i} \bigg)= B \log_2 \bigg( 1+ \sum_{k=1}^K \varrho_{\pi(k),i} - \sum_{k=1}^{i-1} \varrho_{\pi(k),i} \bigg),  \label{eq:Rsum_k}\\
    \overline{R}_{\su}=&\hspace{0.1cm} {\sum\nolimits_{i=1}^K \tau_i R_{\su}^{(i)}}\bigg/\hspace{0.02cm}{\sum\nolimits_{i=1}^K \tau_i} = \sum\nolimits_{i=1}^K \omega_i R_{\su}^{(i)},
    \end{align}
    \noindent where $\varrho_{\pi(k),i}\teq p_{\pi(k),i}\gamma_k(\tv_i) $ can be viewed as the SNR of $D_{\pi(k)}$ during $\tau_i$, $\omega_i= {\tau_i}/{\sum_{i=1}^K \tau_i}$ can be viewed as the weight of $R_{\su}^{(i)}$. And the average total sum rate can be interpreted as the weighted sum rate of each time slot.
    Then, the whole UL transmission process can be viewed as the composition of multiple NOMA-based transmissions with the sum transmission throughput requirement of $\sum_{k=1}^K {L}_k$, i.e., $\sum_{i=1}^K \tau_i R_{\su}^{(i)}=\sum_{k=1}^K {L}_k$.
    From this perspective, the sum transmission delay can be obtained by
    {\setlength\abovedisplayskip{-0.1cm}
    \begin{equation}
		\begin{aligned}
		 \sum\nolimits_{i=1}^K \tau_i & =  {\sum\nolimits_{i=1}^K \tau_i R_{\su}^{(i)}}\bigg/\hspace{0.02cm}{\overline{R}_{\su}} = {\sum\nolimits_{k=1}^K {L}_k}\bigg/\hspace{0.02cm}{\sum\nolimits_{i=1}^K \omega_i R_{\su}^{(i)}}.
		\end{aligned}\label{eq:t_order}
    \end{equation}
    }
    
    \noindent {It can be seen in (\ref{eq:t_order}) that, given the required sum transmission throughput of $\sum_{k=1}^K {L}_k$, to minimize the sum transmission delay, we need to enlarge the denominator, i.e., the average sum rate $\sum_{i=1}^K \omega_i R_{\su}^{(i)}$. For this consideration, on the one hand, we should enlarge $R_{\su}^{(i)}$, which means that we need to minimize $\sum_{k=1}^{i-1} \varrho_{\pi(k),i}$. In other words, we need to rank the devices with smaller SNR $\varrho_{\pi(k),i}$ in the former place. On the other hand, we should enlarge the weight of the sum rate per slot with a larger value.} As indicated by (\ref{eq:Rsum_k}), we have $R_{\su}^{(1)}\geq R_{\su}^{(2)} \geq \cdots\geq R_{\su}^{(K)}$, since more devices will participate in the NOMA-based transmissions in the former time slots. Therefore, we need to increase the weight of those $\tau_i$ with smaller $i$ in the first place. Note that $\tau_i$ is mostly limited by $D_{\pi(i)}$ since $\tau_i \geq \frac{\bar{L}_{\pi(i)}-\sum_{j=1}^{i-1} \tau_j r_{\pi(k),j} }{\log_2 \left[1+{P_{\pi(i)}^{\m}{\gamma}_i(\tv_i)}\right]}$. For example, $\tau_1$ is limited by $D_{\pi(1)}$ since $\tau_1 > \bar{L}_{\pi(1)} / \log_2 \left[1+{P_{\pi(1)}^{\m}{\gamma}_1(\tv_1)} \right]$. Thus, to enlarge the weight of the $\tau_1$, we need to rank those devices with smaller SNR in the former place, which results in ascending order of SNR.

    \tbf{2) Resource Allocation Optimization:}
    After determining the device's order, we focus on the subproblem of resource allocation in this part. First, we have the following proposition.
    \begin{Pro}\label{pro:DK_MPT}
        In the optimal solution to (P1), it follows that  $p^*_{{\pi(K)},i}=P_{\pi(K)}^{\m}$, $\forall i \in \mcal{I}(K)$. 
    \end{Pro}
    
    \begin{myproof}
    If $\exists \hspace{0.1cm} i \in \mcal{I}(K)$ such that $p^*_{{\pi(K)},i}<P_{\pi(K)}^{\m}$, then we can increase the transmit power $p_{{\pi(K)},i}$ to shorten its transmission time without affecting the transmission rates of other devices, since the transmit power of $D_{\pi(K)}$, i.e., $p_{{\pi(K)},i}$, $\forall i \in \mcal{I}(K)$ only influences its own transmission rate $r_{{\pi(K)},i}$ as indicated by (\ref{eq:rate}).
    \end{myproof}
    
    Proposition \ref{pro:DK_MPT}
    indicates that MPT is the optimal solution for the last device $D_{\pi(K)}$ during the entire transmission time slots to minimize its own transmission time consumption under the transmit power constrains. By utilizing Proposition \ref{pro:DK_MPT}, we can directly set $p_{{\pi(K)},i}=P_{\pi(K)}^{\m}$, $\forall i \in \mcal{I}(K)$, which can simplify the variables to be optimized. 
    
    For the optimization w.r.t the other variables in this subproblem, the main difficulty lies in constraint (\ref{eq:con_Lk_P1}). To proceed, we first introduce a new set of variables ${\bm{z}} \teq \{ z_{k,i}= \tau_i \hspace{0.1cm} p_{{\pi(k)},i} \hspace{0.02cm} | \hspace{0.1cm} \forall i \in \mcal{I}(k) , \forall k \in \mcal{K} \}$ and rewrite constraint (\ref{eq:con_Lk_P1}) as
    
    \vspace{-0.3cm}
		\begin{align}
			\sum\nolimits_{i=1}^{k} \tau_i  \log_2 \bigg( 1+ \sum\nolimits_{j=i}^{k}  &\frac{z_{j,i}  \gamma_j(\tv_i)}{\tau_i} \bigg) + 
             \tau_k  \log_2 \bigg( 1+\frac{z_{k,k}  \gamma_k(\tv_k)}{\tau_k}  \bigg)\notag \\ 
		 &	\geq\sum\nolimits_{i=1}^{k-1} \tau_i  \log_2 \bigg( 1+ \sum\nolimits_{j=i}^{k-1}  \frac{z_{j,i}  \gamma_j(\tv_i)}{\tau_i} \bigg) + \bar{L}_k , \hspace{0.2cm} \forall k \in \mcal{K}.\label{eq:SCA_ori}
		\end{align}
    \vspace{-0.3cm}
    
    \noindent Note that (\ref{eq:SCA_ori}) is still not convex due to that the term $ \tau_i  \log_2 \left( 1+ \sum_{j=i}^{k-1}  \frac{z_{j,i}  \gamma_j(\tv_i)}{\tau_i} \right)$ in the right-hand-side of (\ref{eq:SCA_ori}) is concave. However, by utilizing its first-order Taylor expansion at the local points $\hat{\bm{z}}$ and $\hat{\bm{\tau}}$, we can obtain its upper bound as
		\begin{align}
		 	 \tau_i  \log_2 \bigg( 1+ \sum\nolimits_{j=i}^{k-1}  \frac{z_{j,i}  \gamma_j(\tv_i)}{\tau_i} \bigg) \leq	 &\tau_i  \log_2 \bigg( 1+ \sum\nolimits_{j=i}^{k-1}  \frac{\hat{z}_{j,i}  \gamma_j(\tv_i)}{\hat{\tau}_i} \bigg)+ \notag\\
		 & \frac{\sum_{j=i}^{k-1}\bigg(\hat{\tau}_i {z}_{j,i}  \gamma_j(\tv_i)-{\tau}_i \hat{z}_{j,i}  \gamma_j(\tv_i)\bigg)}{\hat{\tau}_i+\sum_{j=i}^{k-1} \hat{z}_{j,i}  \gamma_j(\tv_i)} \teq \mathscr{F}_i(\tau_i,{\bm{z}}).\label{eq:SCA}
		\end{align}
	\noindent Then, the non-convex constraint (\ref{eq:con_Lk_P1}) can be replaced with 
		\begin{align}
				\sum_{i=1}^{k} \tau_i  \log_2 \bigg( 1+  \sum_{j=i}^{k}  &\frac{z_{j,i}  \gamma_j(\tv_i)}{\tau_i} \bigg) + \tau_k  \log_2 \bigg( 1+\frac{z_{k,k}   \gamma_k(\tv_k)}{\tau_k}  \bigg) \geq 
			 \mathscr{F}_i(\tau_i,{\bm{z}}) + \bar{L}_k , \hspace{0.2cm} \forall k \in \mcal{K}.\label{eq:con_SCA}
		\end{align}
	\vspace{-0.4cm}

	\noindent As such, the resource allocation subproblem of (P1) is approximated as the following convex problem
	\begin{subequations}
	\setlength\abovedisplayskip{-0.1cm}
	\setlength\belowdisplayskip{-0.05cm}
    \begin{align}
     (\text{P1.1}): \hspace{0.2cm} \underset{ {\bm{\tau}}, {\bm{z}}  }{\min} \hspace{0.2cm}  & \sum\nolimits_{i=1}^K \tau_{i} \label{eq:con_p_P1.1}\\
      \text { s.t. } \hspace{0.2cm}   &(\ref{eq:con_t_noneg_P1}) ,(\ref{eq:con_SCA}), z_{K,i}=\tau_i  P_{\pi(K)}^{\m},\forall i \in \mcal{K},\\
      &0\leq z_{k,i} \leq \tau_{i} P_{\pi(k)}^{\m}  ,  \forall i \in \mcal{I}(k) , \forall k <K ,   
    \end{align}\label{eq:P1.1}
    \end{subequations}
    
    \noindent which can be successively solved by existing convex optimization solvers such as CVX until convergence is achieved \cite{grant2014cvx}.

    \tbf{3) IRS Beamforming Optimization:}
    To solve the IRS BF optimization subproblem, we have to deal with the sum of the logarithm function and fractions in the constraint (\ref{eq:con_Lk_P1}). To this end, we utilize the FP method to reformulate this constraint \cite{shen2018fractional}. Specifically, by introducing two auxiliary sets of variables ${\bm{\chi}} \teq \{ \chi_{k,i} \hspace{0.05cm} | \hspace{0.1cm} \forall i \in \mcal{I}(k) , \forall k \in \mcal{K} \}$ and ${\bm{\iota}} \teq \{ \iota_{k,i} \hspace{0.05cm} | \hspace{0.1cm} \forall i \in \mcal{I}(k) , \forall k \in \mcal{K} \}$, the constraint (\ref{eq:con_Lk_P1}) is equivalent to
    \begin{equation}
		\begin{aligned}
		 &		\sum\nolimits_{i=1}^{k} u_{k,i}(\tv_i ,\chi_{k,i} ,\iota_{k,i}) \geq {\bar{L}_{\pi(k)}} ,
		\end{aligned}\label{eq:1}
    \end{equation}\noindent
    where 
    
    \vspace{-1.2cm}
		\begin{align}
		  u_{k,i}(\tv_i ,\chi_{k,i} ,\iota_{k,i}) \teq \hspace{0.1cm} & \tau_i   \log_2 \big( 1+ \chi_{k,i} \big) - \tau_i \chi_{k,i} +2\Re \big\{ \iota_{k,i}^H \sqrt{\tau_i ( 1 + \chi_{k,i})p_{{\pi(k)},i}  } \bb_{\pi(k)}^{H} \bv_i \big\}\notag \\
		  &-|\iota_{k,i}|^2 \big( \noi+ \sum\nolimits_{j=i}^{k} p_{{\pi(j)},i} | \bb_{\pi(j)}^{H} \bv_i  |^2  \big)  .
		\end{align}

   \noindent Besides, $\iota_{k,i}$ and $\chi_{k,i}$, $\forall i \in \mcal{I}(k) , \forall k \in \mcal{K}$, are respectively updated by 
   
    \vspace{-0.5cm}

    \begin{align}
    \iota_{k,i}& = \frac{ \sqrt{\tau_i ( 1+ \chi_{k,i})p_{{\pi(k)},i}  } \bb_{\pi(k)}^{H} \bv_i }{ \noi+ \sum_{j=i}^{k} p_{{\pi(j)},i} | \bb_{\pi(j)}^{H} \bv_i  |^2 }  ,\hspace{1cm}
    \chi_{k,i}=
    \begin{cases}\label{eq:iota_chi}
      \frac{ p_{{\pi(k)},i} | \bb_{{\pi(k)}}^{H} \tv_i |^2 }{\sum_{j=i}^{k-1}  p_{{\pi(j)},i} | \bb_{\pi(j)}^{H} \tv_i |^2 + \noi} ,   &1\leq i<k,  \\
     { p_{{\pi(k)},i} | \bb_{\pi(k)}^{H} \tv_i |^2 }/{\noi} , & i=k.
    \end{cases} 
    \end{align}

    \noindent {Since ${\bm{\Psi}}$ is not explicitly expressed in the objective, we convert this subproblem into the following feasibility check problem, whose optimality is shown in \cite{wu2019intelligent}.}
    
    \vspace{-0.4cm}

    \begin{align}
  & \underset{ {\bm{\delta}}, {{\bm{\Psi}}} }{\max} \hspace{0.2cm}   \sum\nolimits_{k=1}^K \delta_{k} \hspace{1cm} \text { s.t. } \hspace{0.2cm}   (\ref{eq:con_v_P1}),	  \sum\nolimits_{i=1}^{k} u_{k,i}(\tv_i ,\chi_{k,i} ,\iota_{k,i}) \geq {\bar{L}_{\pi(k)}}+\delta_{k} , \forall k \in \mcal{K},  \label{eq:P1.2}
    \end{align}
    \vspace{-0.3cm}
    
    \noindent where the slack variables ${\bm{\delta}} \teq \{\delta_1,\cdots,\delta_K\}$ can be interpreted as the ``channel gain residual'' in the optimization, which is beneficial in terms of the convergence manner \cite{wu2019intelligent}.

    \subsection{Proposed Algorithm for (P2)}\label{sec:solu_HMA_en}
    \tbf{1) Device's Ordering Principle:}
    Under the transmit energy constraints, for a special case of a two-device system without IRS,  the descending order based on the SNR in the TDMA scheme will undermine the performance of the proposed I-HMA protocol for the special case with the same UL transmission throughput requirement and energy constraint for each device, since it always leads to $E_{\pi(k)} \leq E_{\pi(k)}^{\td}$, $\forall k >1$ \cite{zeng2019delay}. However, with the assistance of the IRS, the equivalent channels vary with different IRS BF patterns, which may change the optimal device's rearranged order and thus it is difficult to evaluate the specific performance under different rearranged orders. Nevertheless, given ${\bm{\Psi}}={\bm{\Psi}}^{\td}$, by examining the KKT conditions as given in Appendix D, if we rearrange the devices according to the ascending order of TDMA-based SNR, we always have $E_{\pi(k)}> E_{\pi(k)}^{\td}({\bm{\Psi}})$, $\forall k>1$, which can guarantee the superiority of the proposed I-HMA protocol over the I-TDMA protocol. Therefore, in this case, we still adopt the ascending order of TDMA-based SNR as the ordering principle. 
    The superiority of the proposed device's ordering is also validated by the simulation results as will be presented in Section \ref{sec:simu}.

    \tbf{2) Resource Allocation Optimization:}
	Similar to the manipulations for (P1), the solution of the resource allocation subproblem for (P2) can be obtained by solving the following convex problems 
    
    \vspace{-0.2cm}
    \begin{align}
    (\text{P2.1}): \hspace{0.2cm} &\underset{ {\bm{\tau}}, {\bm{z}}  }{\min} \hspace{0.2cm}   \sum\nolimits_{i=1}^K \tau_{i}   \hspace{1cm} \text { s.t. } \hspace{0.2cm}   (\ref{eq:con_t_noneg_P1}) ,(\ref{eq:con_SCA}),  0\leq z_{k,i} \leq E_{\pi(k)}  ,  \forall i \in \mcal{I}(k) , \forall k \in \mcal{K}.   
    \end{align}

    \tbf{3) IRS Beamforming Optimization:}
    For the subproblem of IRS BF optimization regarding to (P2), the aforementioned transformation for (P1) is also applicable, which can be tackled by solving the feasibility check problem as given by (\ref{eq:P1.2}).

    \subsection{Solution for I-TDMA Based Optimization}\label{sec:solu_TDMA}
    
    \tbf{1) Transmit Power Limited Case: }
     It is easy to verify that the optimal $ {\bm{p}}^{\td ^*}$, ${\bm{\tau}}^{\td ^*}$ and ${\bm{\Psi}}^{\td ^*}$ can be obtained as
    \begin{subequations}
    \setlength\abovedisplayskip{-0.05cm}
    \begin{numcases}{}
     p_{{\pi(k)}}^{\td^*}=P_{\pi(k)}^{\m},\forall k \in \mcal{K},\\
     \tau_{k}^{\td ^*}=\bar{L}_{\pi(k)}/\log_2 \big[1+p_{{\pi(k)}}^{\td ^*}\gamma_k(\tv_k^{\td ^*} )\big]  ,\forall k \in \mcal{K},\\
      \big[\tv_k^{\td ^*}\big]_n=e^{j\angle{\left[\bb_{\pi(k)}\right]_n}},\forall n \in \mcal{N},\forall k \in \mcal{K}.\label{eq:TDMA_vk}
    \end{numcases}\label{eq:opt_TDMA_po}
    \end{subequations}
    \vspace{-0.2cm}

    \tbf{2) Transmit Energy Limited Case:}
    Based on the discussion in Appendix D, the optimal ${\bm{\tau}}^{\td ^*}$ and $ {\bm{p}}^{\td ^*}$ are calculated by
    \begin{subequations}
    \setlength\abovedisplayskip{-0.05cm}
    \begin{numcases}{}
     \tau_{k}^{\td ^*}=\mathscr{G}_k(\tv_k^{\td ^*}), \hspace{0.4cm}\forall k \in \mcal{K},\\
      p_{{\pi(k)}}^{\td^*}=E_{\pi(k)}/\tau_{k}^{\td ^*},\hspace{0.2cm}\forall k \in \mcal{K},
    \end{numcases}\label{eq:opt_TDMA_en}
    \end{subequations}\vspace{-0.2cm}
    
    \noindent where $\mathscr{G}_k(\tv_k)$ is given by (\ref{eq:TDMA_E}) and the optimal IRS BF patterns set ${\bm{\Psi}}^{\td ^*}$ is given by (\ref{eq:TDMA_vk}).

    \subsection{Solution for I-NOMA Based Optimization}\label{sec:solu_NOMA}

    \tbf{1) Transmit Power Limited Case:}
    The optimal $ {\bm{p}}^{\no ^*}$, ${\bm{\tau}}^{\no ^*}$ and ${\bm{\Psi}}^{\no ^*}$ in this scenario can be obtained by solving the following problem
    \begin{subequations}
    \setlength\abovedisplayskip{-0.01cm}
    \setlength\belowdisplayskip{-0.01cm}
    \begin{align}
    (\text{P1-N}): \hspace{0.2cm} \underset{ {\tau_1}, \{p_{k,1}\} ,\tv_1 }{\min} \hspace{0.2cm} &   \tau_{1}  \label{eq:obj_P3}\\
    \text { s.t. } \hspace{0.6cm}   & \tau_1 r_{{\pi(k)},1} \geq L_{\pi(k)}, \forall k \in \mcal{K},\label{eq:con_Lk_P3} \\
    & 0\leq p_{{\pi(k)},1} \leq P_{\pi(k)}^{\m},\forall k \in \mcal{K},   \label{eq:con_p_P3} \\
    & \tau_{1} \geq 0, \left|\left[ \tv_1 \right]_{n}\right|=1,\forall n \in \mcal{N}, \label{eq:con_v_P3}
    \end{align}
    \end{subequations}
    \noindent which can be solved via the manipulations similar to (P1).
    
    \begin{algorithm}[!htbp]
    \caption{Proposed Algorithm for (P1) and (P2).}\label{alg:AO}
    \begin{algorithmic}[1]\setstretch{1.3}
    \STATE Initialize the device's ordering as {\small$\pi(k)=k,\forall k \in \mcal{K}$}.
    \STATE Obtain {\small${\bm{\Psi}}^{\td^*}$} by (\ref{eq:TDMA_vk}), obtain {\small${\bm{\tau}}^{\td ^*}$} and {\small$ {\bm{p}}^{\td ^*}$} by (\ref{eq:opt_TDMA_po}) for (P1), or by (\ref{eq:opt_TDMA_en}) for (P2).
    \STATE Calculate the TDMA-based SNR {\small${\bm \varrho}^{\td}$} with (\ref{eq:SNR_P}). 
    \STATE Update the order {\small${\bm{\pi}}$} such that {\small$\varrho_{\pi(1)}^{\td} \leq  \cdots \leq \varrho_{\pi(K)}^{\td}$}.
    \STATE Obtain the NOMA-based solution {\small$\mathscr{S}^{\no ^*}({\bm{\Psi}}^{\no ^*})$} by solving (P1-N) or (P2-N).  
    \STATE Compute {\small${L}_{\pi(k)}^{{\no}}( \tv_1^{\no^*}) ,\forall k>1$}, for (P1); or compute  {\small${E}_{\pi(k)}^{{\no}}(\tv_1^{\no^*})$} and {\small${E}_{\pi(k)}^{{\td}}({\bm{\Psi}}^{\td^*})$}, {\small$\forall k>1$}, for (P2).
    \IF{{\small${L}_{\pi(k)} \leq   {L}_{\pi(k)}^{{\no}}( \tv_1^{\no^*}) ,\forall k>1$} for (P1); or {\small${E}_{\pi(k)} \geq  {E}_{\pi(k)}^{{\no}}(\tv_1^{\no^*}), \forall k >1$}, for (P2),}
    \STATE Set {\small${\bm{\tau}}^{\hy^*}={\bm{\tau}}^{\no^*}$}, {\small${\bm{p}}^{\hy^*}={\bm{p}}^{\no^*}$}, and {\small${\bm{\Psi}}^{\hy^*}={\bm{\Psi}}^{\no^*}$}.
    \ELSIF{{\small${E}_{\pi(k)} \leq  {E}_{\pi(k)}^{{\td}}({\bm{\Psi}}^{\td^*}), \forall k >1$}, for (P2),}
    \STATE Set {\small${\bm{\tau}}^{\hy^*}={\bm{\tau}}^{\td^*}$}, {\small${\bm{p}}^{\hy^*}={\bm{p}}^{\td^*}$}, and {\small${\bm{\Psi}}^{\hy^*}={\bm{\Psi}}^{\td^*}$}.
    \ELSE
    \STATE Initialize the iteration index {\small$n$} to be 0.
    \STATE Initialize {\small$\tau^{(n)}_{\pi(k)}=\tau_{\pi(k)}^{\td^*}$}, {\small$p^{(n)}_{{\pi(k)},k}=p_{\pi(k)}^{\td^*}$}, {\small$p^{(n)}_{{\pi(k)},i}=0$}, {\small$\forall i \in \mcal{I}(k-1) $}, and {\small$\forall k \in \mathcal{K}$}.
    \REPEAT
    \REPEAT
    \STATE Update {\small${\bm{\iota}}$} and {\small${\bm{\chi}}$} with the given {\small${\bm{\tau}}^{(n)}$}, {\small${\bm{p}}^{(n)}$} and {\small${\bm{\Psi}}^{(n)}$} by (\ref{eq:iota_chi}).
    \STATE Update {\small${\bm{\Psi}}^{(n)}$} by solving problem (\ref{eq:P1.2}) with the given {\small${\bm{\tau}}^{(n)}$}, {\small${\bm{p}}^{(n)}$}, {\small${\bm{\iota}}$} and {\small${\bm{\chi}}$}.
    \UNTIL {problem (\ref{eq:P1.2}) converges.}
    \STATE Set {\small${\bm{\Psi}}^{(n+1)}={\bm{\Psi}}^{(n)}$}
    \STATE Calculate {\small${\bm{z}}^{(n)}$} according to {\small${\bm{p}}^{(n)}$} and {\small${\bm{\tau}}^{(n)}$}. 
    \STATE Obtain {\small${\bm{\tau}}^{(n+1)}$} and {\small${\bm{p}}^{(n+1)}$} by solving problem (P1.1) for (P1), or (P2.1) for (P2), at the local point {\small${\bm{z}}^{(n)}$} and {\small${\bm{\tau}}^{(n)}$} with the given {\small${\bm{\Psi}}^{(n+1)}$}.
    \STATE Set {\small$n=n+1$}.
    \UNTIL {the fractional decrease of the sum transmission delay is below a predefined threshold.}
    \STATE Set {\small${\bm{\tau}}^{\hy^*}={\bm{\tau}}^{(n)}$}, {\small${\bm{p}}^{\hy^*}={\bm{p}}^{(n)}$}, and {\small${\bm{\Psi}}^{\hy^*}={\bm{\Psi}}^{(n)}$}.
    \ENDIF
    \STATE \tbf{Output:} {\small${\bm{\pi}}$}, {\small${\bm{\tau}}^{\hy^*}$}, {\small${\bm{p}}^{\hy^*}$} and {\small${\bm{\Psi}}^{\hy^*}$}.
    \end{algorithmic}
    \end{algorithm}
    \setlength{\textfloatsep}{10pt}
    
    \tbf{2) Transmit Energy Limited Case:}
    Likewise, by applying similar manipulations for (P2), the optimal $ {\bm{p}}^{\no ^*}$, ${\bm{\tau}}^{\no ^*}$ and ${\bm{\Psi}}^{\no ^*}$ can be obtained  by solving the following problem
    \begin{align}
    \setlength\abovedisplayskip{-0.01cm}
    \setlength\belowdisplayskip{-0.01cm}
    \hspace{-0.3cm}(\text{P2-N}):   \hspace{0.2cm}&\underset{ {\tau_1}, \{p_{k,1}\} ,\tv_1 }{\min} \hspace{0cm}   \tau_{1}  \hspace{1cm}\text { s.t. } \hspace{0cm}    (\ref{eq:con_Lk_P3}),(\ref{eq:con_v_P3}),  0\leq p_{{\pi(k)},1}\tau_1 \leq E_{\pi(k)},\forall k \in \mcal{K}.  \label{eq:con_p_P4}
    \end{align}
    
    \vspace{-0.5cm}

    \subsection{Computational Complexity Analysis}
    To sum up, the solution of (P1) and (P2) can be obtained as summarized in Algorithm \ref{alg:AO}.
    The computational complexity of (P1) and (P2) is given by $\mcal{O}\big(I_{\text{AO}}\big[(K+{K^2}/{4})^{3.5}+I_{\text{FP}}(KN)^{3.5}\big]\big)$, whereas the computational complexity of (P1-N) and (P2-N) is given by $\mcal{O}\left(I_{\text{AO}}\left[K^{3.5}+I_{\text{FP}}N^{3.5}\right]\right)$ based on the analytical results in \cite{wang2014outage}. Since the optimal solution of TDMA-based optimization is given by the  (\ref{eq:TDMA_vk}), (\ref{eq:opt_TDMA_po}) and (\ref{eq:opt_TDMA_en}), its computational complexity computational complexity is caused by the calculation of $\bb_{{\pi(k)}}^{H} \tv$ in $\gamma_k(\tv)$, which is given by $\mcal{O}(KN)$. As such, the computational complexity can be reduced greatly by resorting to the special cases of TDMA-based optimization problem or NOMA-based optimization problem with the proposed algorithm if the condition as indicated by Propositions \ref{pro:HMA_eq_NOMA} and \ref{pro:bound} holds.

    \vspace{-0.2cm}
	\section{Simulation Results}\label{sec:simu}
    
\begin{figure}[!t]
\flushright
\begin{minipage}[t]{0.39\textwidth}
\centering
\setlength{\abovecaptionskip}{-0.1cm}
		\subfigure{
		\includegraphics[height=0.39\textwidth]{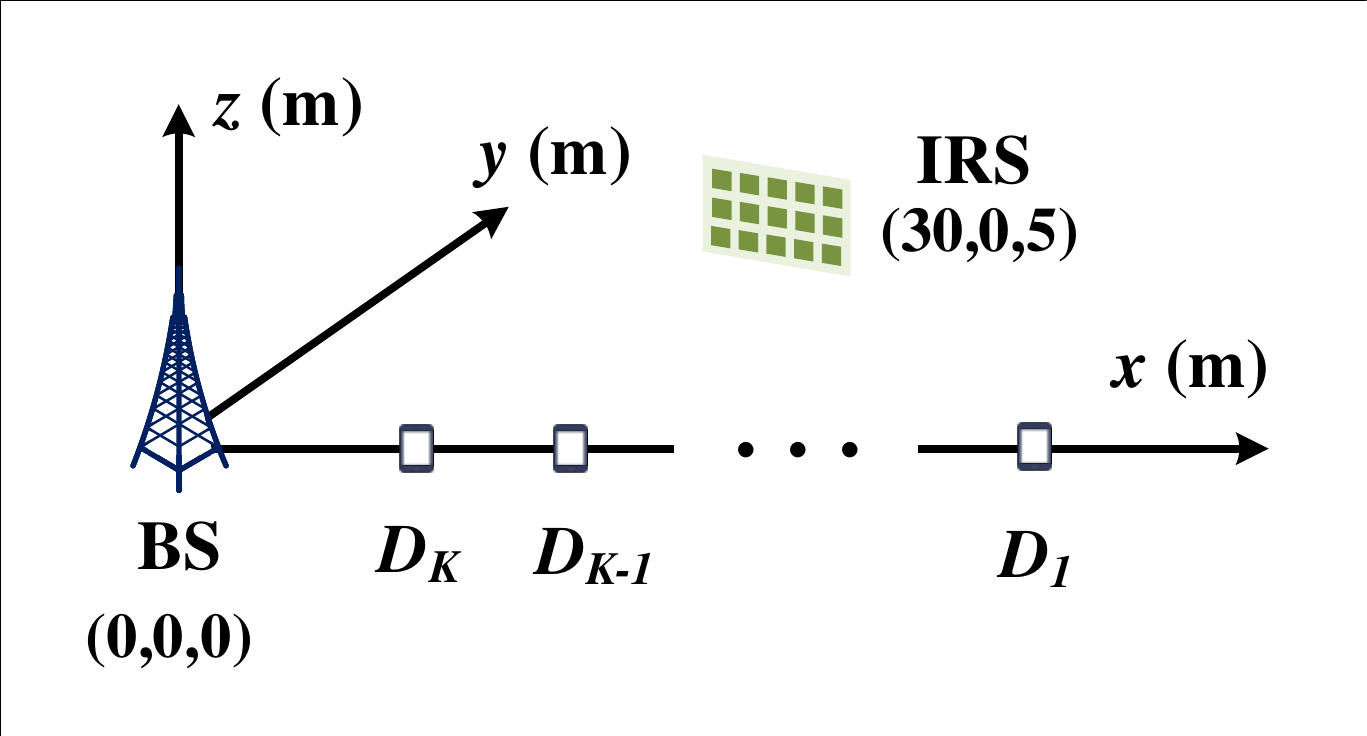}}
		\caption{The deployment of the IRS-aided  UL \newline transmission system.}
\label{fig:simu}
\end{minipage}
\begin{minipage}[t]{0.61\textwidth}
\centering
\subfigure[Asymmetric distributed devices.]{
			\label{fig:asy_dep}
			\includegraphics[height=0.21\textwidth]{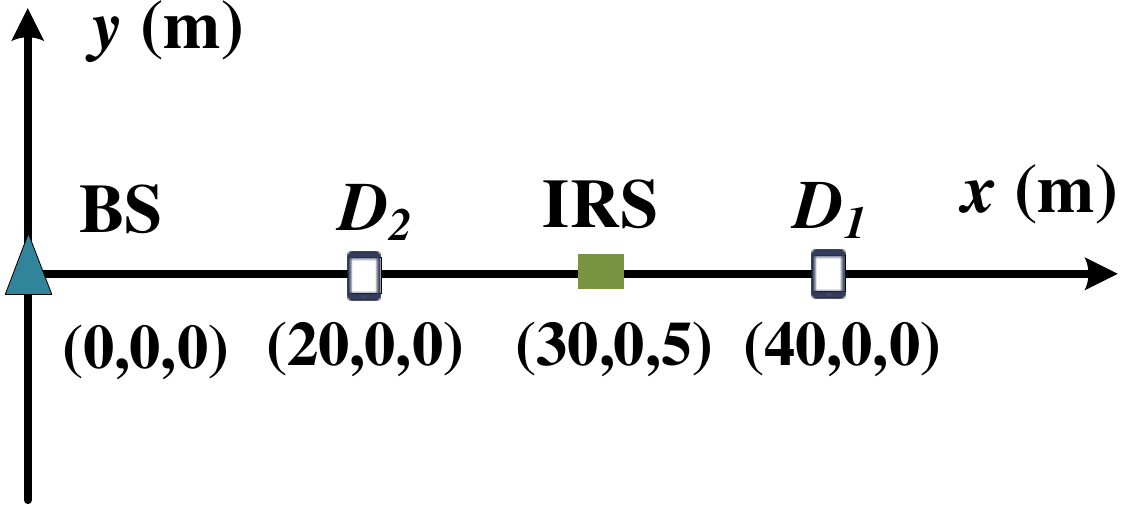}}
		\subfigure[Symmetric distributed devices.]{
			\label{fig:sy_dep}
			\includegraphics[height=0.21\textwidth]{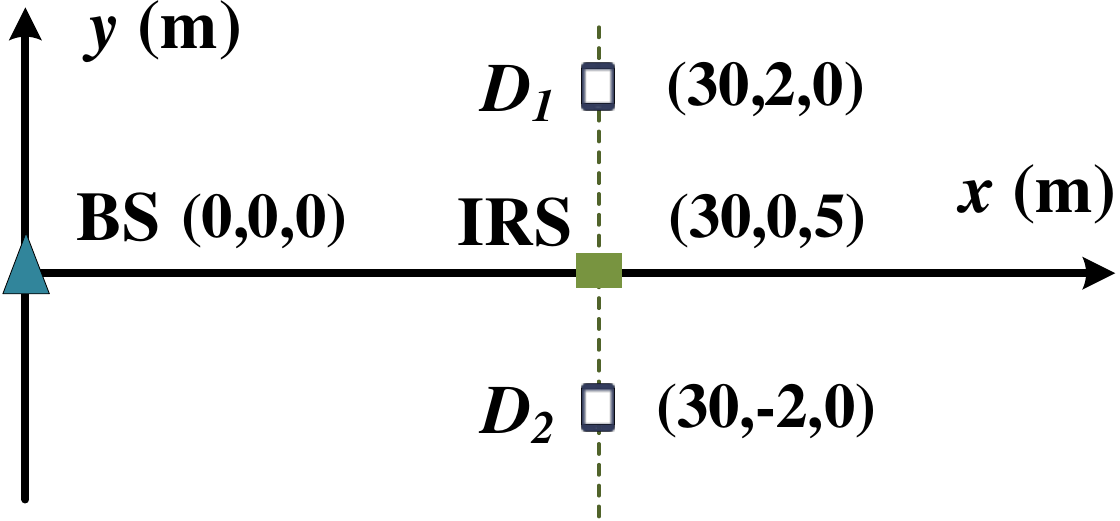}}
			\setlength{\abovecaptionskip} {0.05cm}
			\caption{The top view of the deployment of the two-device system.} 
\label{fig:loca}
\end{minipage}
\end{figure}

    In this section, simulation results are presented to numerically validate our analytical propositions and the effectiveness of the proposed I-HMA protocol in the IRS-aided UL transmission system.
    As shown in Fig. \ref{fig:simu}, the BS and IRS are located at $(0, 0, 0)$ meter (m) and $(30,0,5)$ m, respectively, and the devices are distributed along the $x$-axis, if not specified otherwise. 
    For the involved channels\footnote{Note that no particular assumption on the channel models and device's location has been made in our system model and the problem formulation shown in Section II. Therefore, all the analytical results regarding performance comparison of the proposed I-HMA protocol with the I-TDMA and I-NOMA schemes in IRS-aided energy-constrained UL communication systems and algorithmic design are applicable for any channel model and device's location.}, we adopt the path loss model in \cite{wu2019intelligent}, where the path loss exponent is set to be 2.2 for $\bg$ and $\bh_{r,k}$, whereas 3.6 for $h_{d,k}$, $\forall k\in \mcal{K}$. Rayleigh fading is adopted to characterize the small-scale fading for all the channels \cite{wu2021irs}. Unless otherwise stated, the parameters are set as follows: $B=500$ kHz, $\noi=-80$ dBm and $N=50$ \cite{wu2021irs}. Other simulation parameters given in the following are set according to \cite{ding2018delay,zeng2019delay,zhu2020resource,ding2022hybrid}.

    The annotations for the curves in the simulation results are given as follows. 1) \textbf{HMA:} the approach in Sec. \ref{sec:solu_HMA_po} or Sec. \ref{sec:solu_HMA_en}; 2) \textbf{TDMA:} the approach in Sec. \ref{sec:solu_TDMA}; 3) \textbf{NOMA:} the approach in Sec. \ref{sec:solu_NOMA}; 4) \textbf{opt-o:} find the optimal device's order by exhausting all possible orders; 5) \textbf{pro-o:} employing the proposed principle in Sec. \ref{sec:solu}; 6) \textbf{des-o:} rearrange devices according to the descending order of TDMA-based SNR; 7) \textbf{d-IRS:} utilizing dynamic IRS BF and optimizing the IRS BF patterns with the proposed method in Sec. \ref{sec:solu_HMA_po}3; 8) \textbf{s-IRS:} utilizing static IRS BF and optimizing the IRS BF vector with the proposed method in Sec. \ref{sec:solu_HMA_po}3 by replacing $\tv_k$ with $\tv_1$; 9) \textbf{w/o IRS:} without the assistance of IRS.

    \vspace{-0.2cm}

    \subsection{Comparison Among Different MA Protocols}
    \begin{figure}
		\centering  
		\subfigbottomskip=2pt 
		\subfigcapskip=-5pt 
		\subfigure[Asymmetric distributed devices (${L}_2^{{\no}}=68.39$ Kbits).]{
			\label{fig:P2_asy}
			\includegraphics[height=0.26\textwidth]{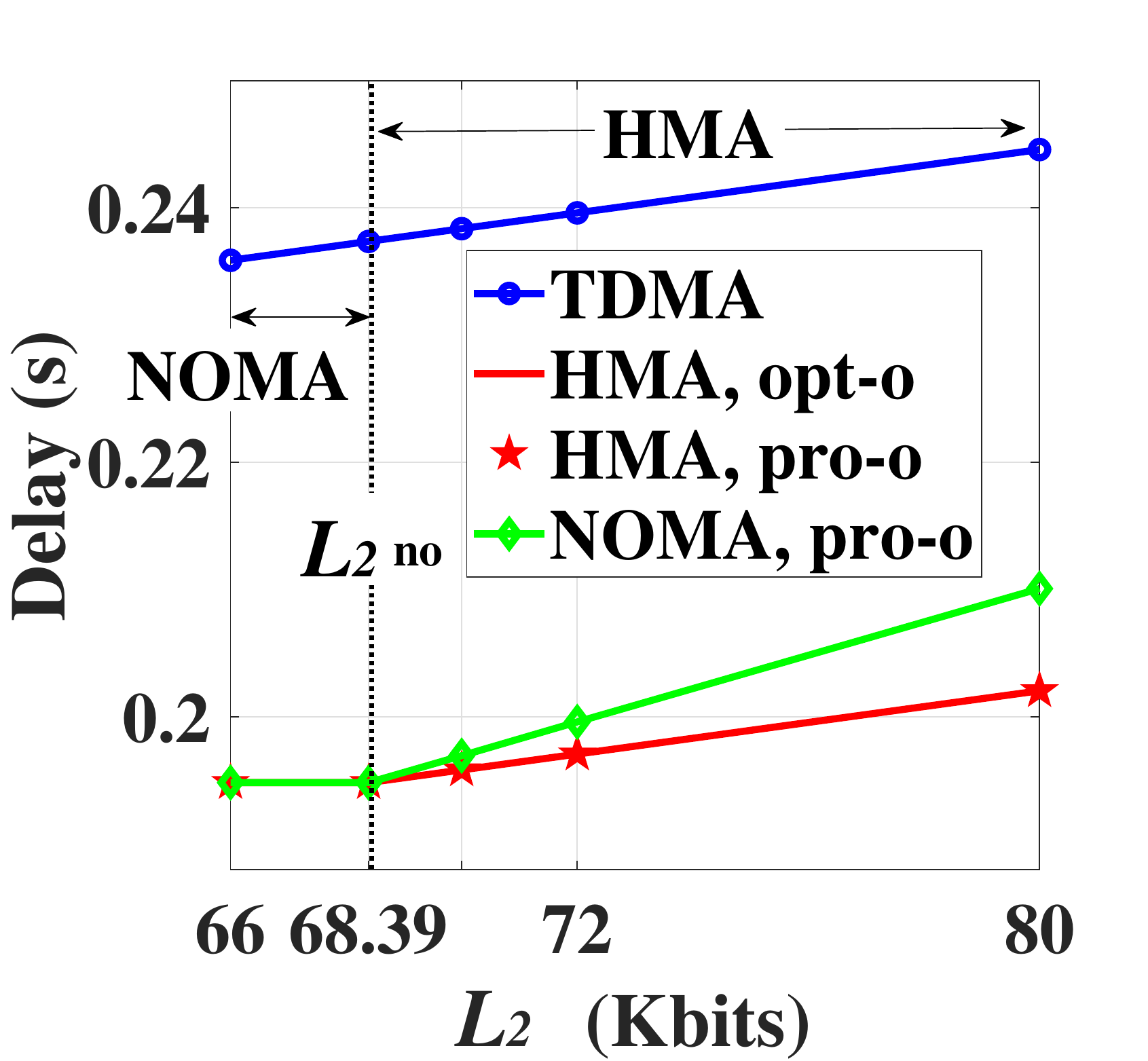}}
		\hspace{1cm}
		\subfigure[Symmetric distributed devices (${L}_2^{{\no}}=3.15$ Kbits).]{
			\label{fig:P2_sy}
			\includegraphics[height=0.26\textwidth]{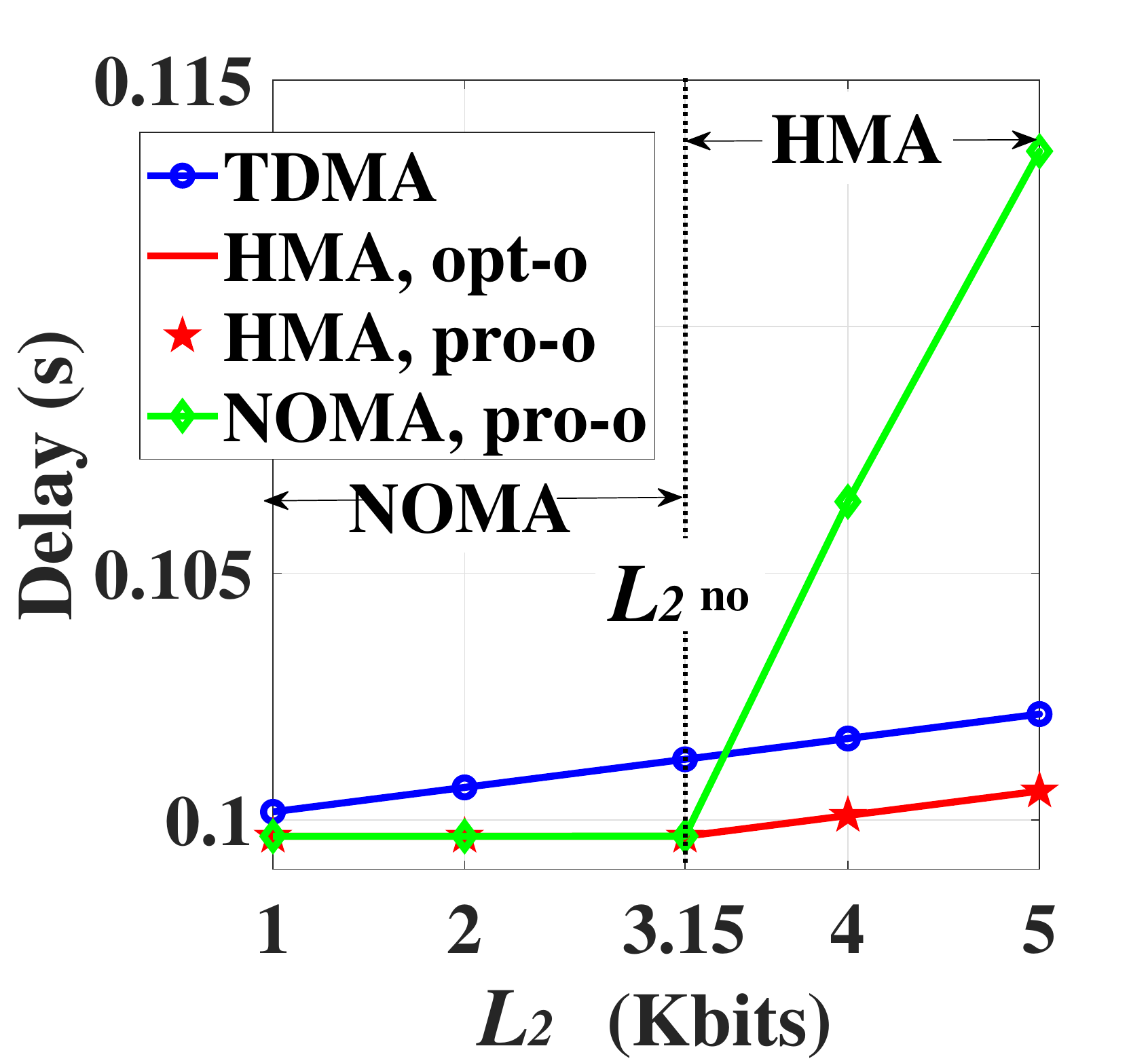}}
		\caption{Sum transmission delay under the transmit power constraints ($K=2$, $P_1^{\m}=P_2^{\m}=5$ dBm, $L_1=200$ Kbits).} 
		\label{fig:P2}
	\end{figure}

	\begin{figure}
		\centering  
		\subfigbottomskip=2pt 
		\subfigcapskip=-5pt 
		\subfigure[Asymmetric distributed devices ($E_2^{\td}=0.01$ J, $E_2^{\no}=0.084$ J).]{
			\label{fig:E2_asy}
			\includegraphics[height=0.26\textwidth]{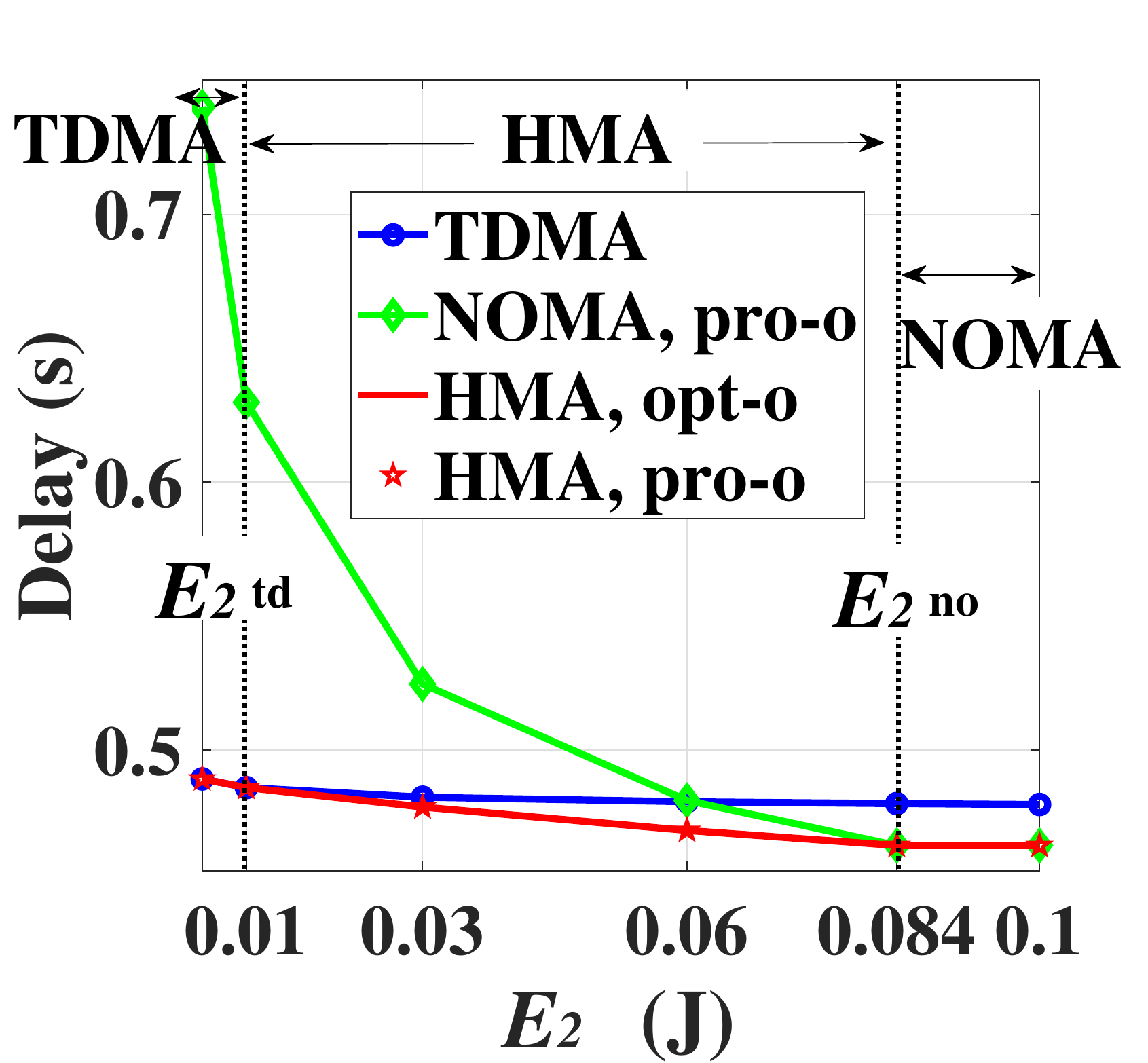}}
		\hspace{1cm}
		\subfigure[Symmetric distributed devices ($E_2^{\td}=0.147$ J, $E_2^{\no}=2.156$ J).]{
			\label{fig:E2_sy}
			\includegraphics[height=0.26\textwidth]{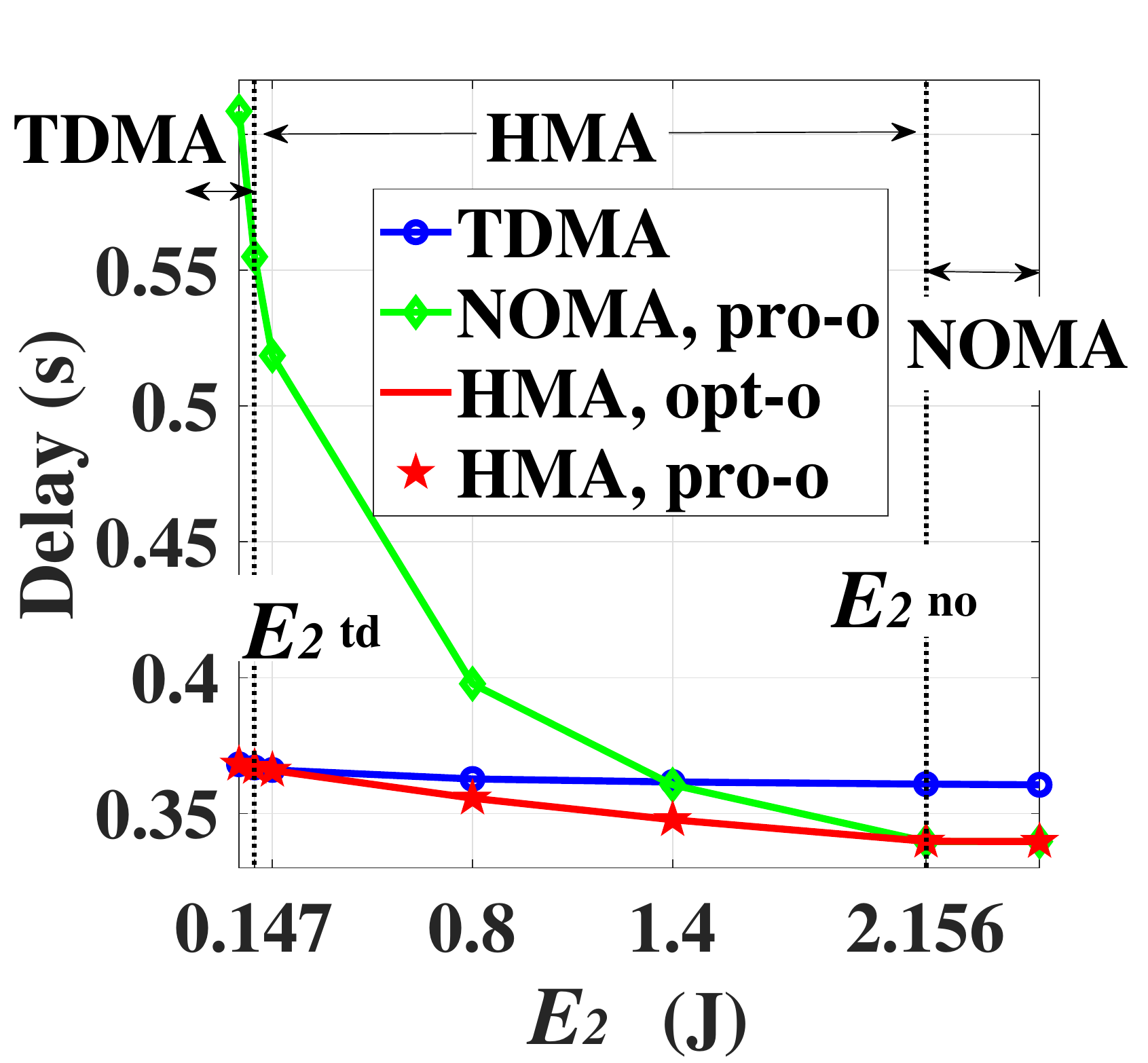}}
		\caption{Sum transmission delay under the transmit energy constraints ($K=2$, $E_1=0.1$ J, $L_1=2000$ Kbits, $L_2=200$ Kbits).} 
		\label{fig:E2}
	\end{figure}
	In Figs. \ref{fig:P2} and \ref{fig:E2}, we investigate the performance of different MA protocols under the transmit power constraints and the transmit energy constraints, respectively. Two kinds of deployment, namely the asymmetric distribution and symmetric distribution are considered as shown in Fig. \ref{fig:loca} with $K=2$. Specifically, for asymmetric deployment as shown in Fig. \ref{fig:asy_dep}, two devices have different distances to the BS, which are located at $(20,0,0)$ m and $(40,0,0)$ m, respectively. In contrast, for symmetric deployment as shown in Fig. \ref{fig:sy_dep}, two devices have equal distance to the BS, which are located at $(30,2,0)$ m and $(30,-2,0)$ m, respectively. 

    {It can be seen in Fig. \ref{fig:P2} that, for a transmit power limited system, the proposed I-HMA protocol always outperforms the I-TDMA protocol for both deployments, as proved by Proposition \ref{pro:HMA_sup}. Note that the performance gap between the proposed I-HMA protocol with the I-TDMA protocol is more significant when the devices are distributed asymmetrically compared with its symmetrical counterpart. This is because NOMA can better reduce transmission delay under asymmetric device's channels, which enables the proposed I-HMA protocol to reap more benefits from the NOMA protocol.
    For the comparison between the proposed I-HMA protocol with the I-NOMA protocol, we calculate ${L}_2^{{\no}}$ by utilizing (\ref{eq:L_bound}). It is observed that when ${L}_2\leq{L}_2^{{\no}}$, the performance of the proposed I-HMA protocol is the same as that of the I-NOMA protocol, whereas the proposed I-HMA protocol outperforms the I-NOMA protocol when ${L}_2>{L}_2^{{\no}}$, which validates the accuracy of Propositions \ref{pro:HMA_eq_NOMA} and \ref{pro:HMA_sup_no}.
    Moreover, we also compare the performance of the proposed I-HMA protocol under the proposed ordering principle (\textbf{``HMA, pro-o''}) and the optimal order by exhausting all possible orders (\textbf{``HMA, opt-o''}). In particular, the performance of the \textbf{``HMA, pro-o''} protocol is in accordance with that of the \textbf{``HMA, opt-o''} protocol, which indicates the effectiveness of the proposed ordering principle.}

	\begin{figure}[t]
		\centering
		\subfigure[The sum transmission delay versus the path loss exponent of the cascaded links ($K=3$, $P_1^{\m}=P_2^{\m}=P_3^{\m}=5$ dBm, $L_1=L_2=L_3=10$ Kbits).]{
			\label{fig:alpha_IRS}
			\includegraphics[height=0.33\textwidth]{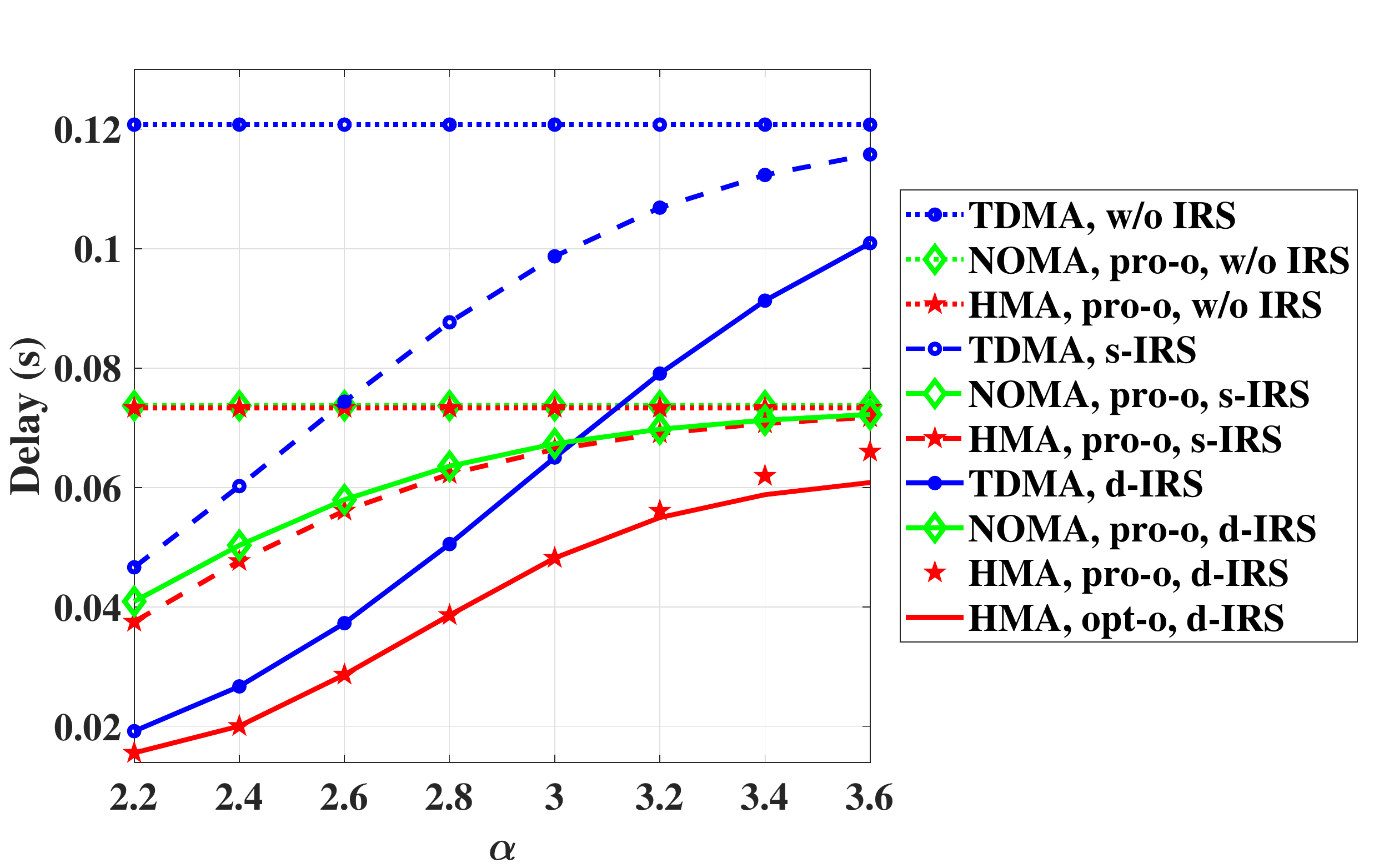}}
			\hspace{0.5cm}
		\subfigure[The sum transmission delay versus the number of reflecting elements ($K=2$, $P_1^{\m}=P_2^{\m}=5$ dBm, $L_1=20$ Kbits, $L_2=40$ Kbits).]{
			\label{fig:N}
			\includegraphics[height=0.33\textwidth]{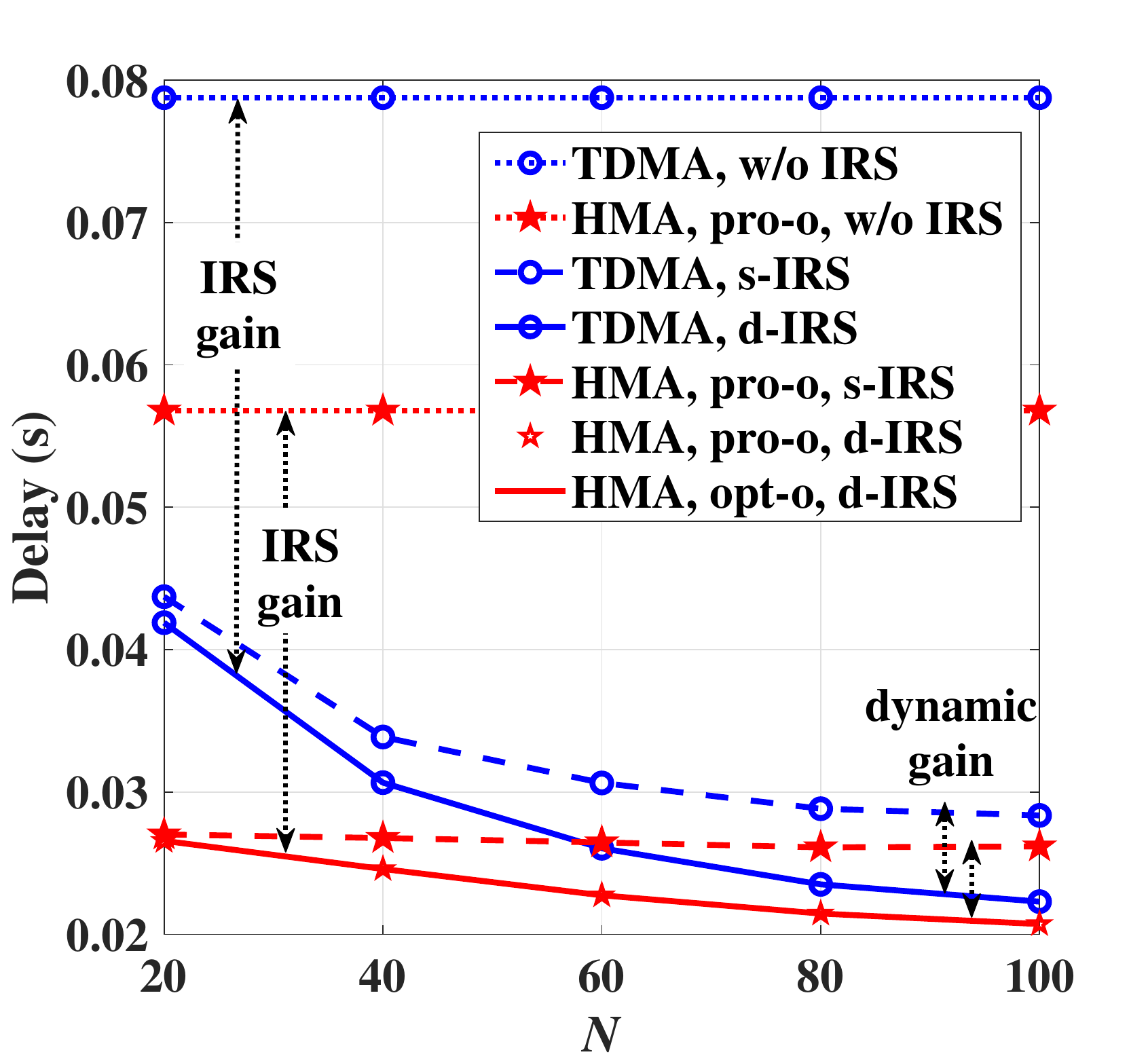}}
			\caption{The effect of IRS on the sum transmission delay under the transmit power constraints.}
		\label{fig:IRS}
	\end{figure} 

	On the other hand, for a transmit energy limited system as shown in Fig. \ref{fig:E2}, the performance of the proposed I-HMA protocol outperforms the other two benchmarks (I-TDMA and I-NOMA) when ${E}_2^{{\no}}>  {E}_2>  {E}_2^{{\td}}$, as indicated by Proposition \ref{pro:bound}. By utilizing (\ref{eq:E_no}) and (\ref{eq:E_td}), we calculate the value of $E_2^{\no}$ and $E_2^{\td}$, which are exactly consistent with the simulation results. One can observe that when ${E}_2\leq  {E}_2^{{\td}}$, the proposed I-HMA protocol degrades to the I-TDMA protocol, whereas the proposed I-HMA protocol reduces to the I-NOMA protocol when ${E}_2\geq {E}_2^{{\no}}$, which verifies the correctness of Proposition \ref{pro:bound}. Besides, it is noted that when the devices are distributed asymmetrically, the value of ${E}_2^{{\no}}$ is smaller than its symmetrical counterpart, which indicates that the NOMA protocol plays a more significant role under the asymmetric deployment because the
	the proposed I-HMA scheme is equivalent to the I-NOMA scheme for a wider range in the former scenario.

    \subsection{The Effect of IRS}

    In Fig. \ref{fig:IRS}, we evaluate the advantages of introducing the IRS into the system. Specifically, in Fig. \ref{fig:alpha_IRS}, we plot the sum transmission delay versus the path loss exponent (denoted by $\alpha$) of the IRS-involved links, i.e., $\bg$ and $\bh_{r,k}$, $\forall k \in \mcal{K}$ in a power limited system, with the parameters set as $K=3$, $D_1=(40,0,0)$ m, $D_2=(30,0,0)$ m, $D_3=(20,0,0)$ m, $P_1^{\m}=P_2^{\m}=P_3^{\m}=5$ dBm, and $L_1=L_2=L_3=10$ Kbits. Note that the path loss exponent of the IRS-involved links can characterize the weight of the role the IRS plays in the system. In particular, the smaller the path loss exponent of the IRS-involved links is, the more prominent the effect of those links will be. It is found that a significant performance gain is achieved after introducing the IRS into the system as compared to the case without IRS for all MA schemes. And this performance gap is more significant as $\alpha$ decreases, which indicates that the IRS does play an important role in improving the system performance. 
    Besides, we also examine the benefit of dynamic IRS BF. Specifically, a significant performance gain can be achieved by utilizing the dynamic IRS BF compared with the performance of static IRS BF.
    And when the proposed I-HMA protocol can only use static IRS BF, the performance gain compared with the I-NOMA protocol will be very small. This is expected since the static IRS BF setting results in a less flexible channel reconfiguration. By comparing the performance of \textbf{``HMA, d-IRS''} with \textbf{``TDMA, d-IRS''} and \textbf{``HMA, s-IRS''}, {it can be verified that the proposed dynamic I-HMA protocol not only can reap the benefit of the time-saving advantage of the NOMA protocol compared with the TDMA protocol, but can also get the dynamic BF gain of the TDMA protocol compared with the NOMA protocol, which indicates the effectiveness of the proposed design.} In addition, it is also observed that the proposed ordering principle performs well compared with the optimal order, which provides a useful guideline to rearrange the devices in practice.

    To show the effect of the number of reflecting elements at the IRS on the system performance, we plot the sum transmission delay versus $N$ in Fig. \ref{fig:N}, with the parameters set as $K=2$, $D_1=(40,0,0)$ m, $D_2=(20,0,0)$ m, $P_1^{\m}=P_2^{\m}=5$ dBm, $L_1=20$ Kbits, and $L_2=40$ Kbits. It can be seen that the sum transmission delay declines significantly with the assistance of the IRS under both MA protocols. More importantly, {the proposed design significantly reduces the sum transmission delay even with a very small number of reflecting elements at IRS, which indicates the great potential of utilizing IRS.} Besides, we can also observe an additional performance gain by shifting the IRS BF patterns over time compared with the static IRS BF scheme, with the performance gap increasing as $N$ increases, which demonstrates the effectiveness of the dynamic IRS BF design.

    \begin{figure}
		\centering
		\subfigure[The sum transmission delay under the transmit power constraints ($P_k^{\m}=10$ dBm, $L_k=10 (K-k+1)$ Kbits).]{
			\label{fig:order_p}
			\includegraphics[width=0.42\textwidth]{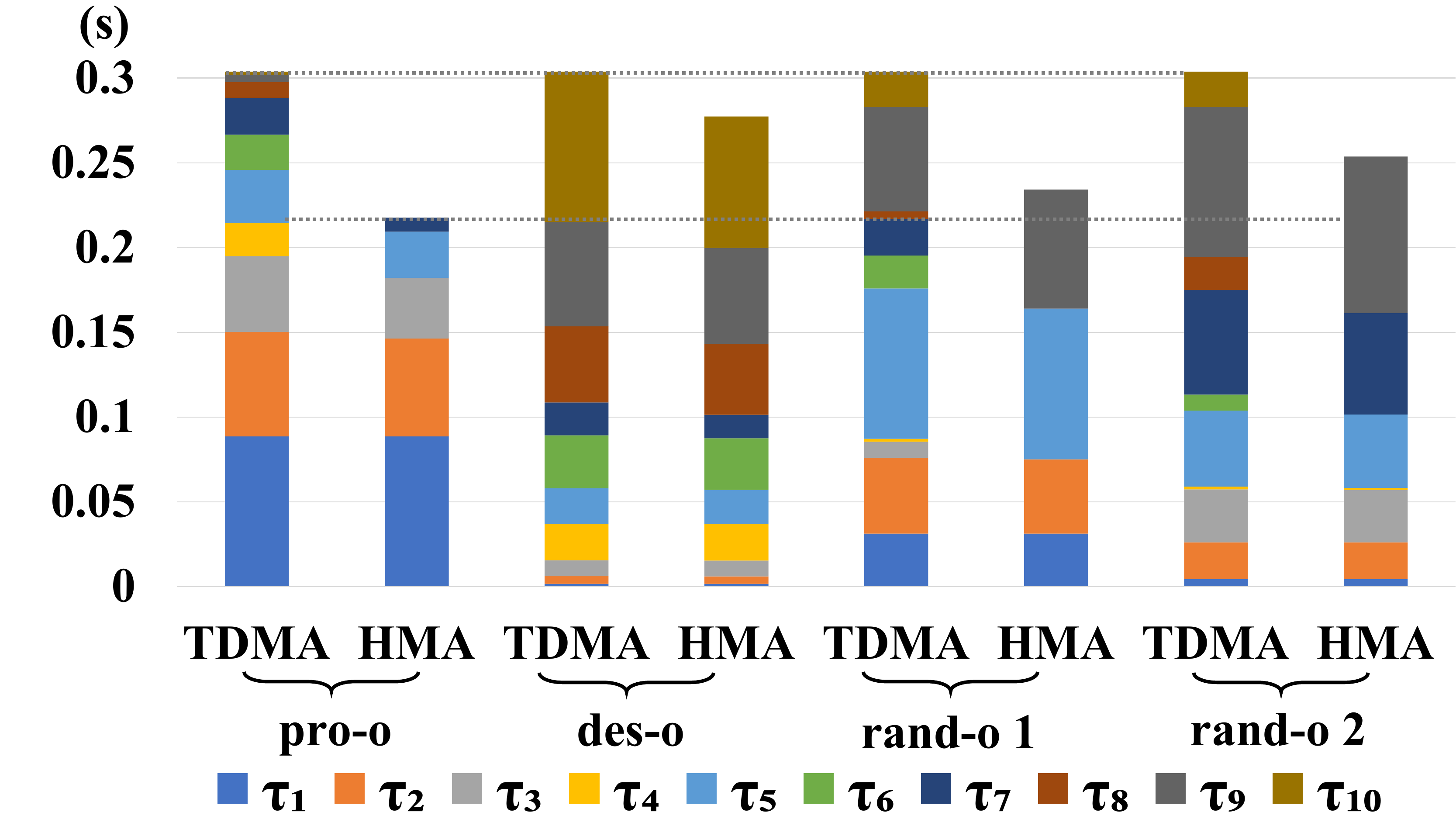 }}\hspace{1.5cm}
		\subfigure[The sum transmission delay under the transmit energy constraints ($E_k=k$ J, $L_k=200 (K-k+1)$ Kbits).]{
			\label{fig:order_e}
			\includegraphics[width=0.42\textwidth]{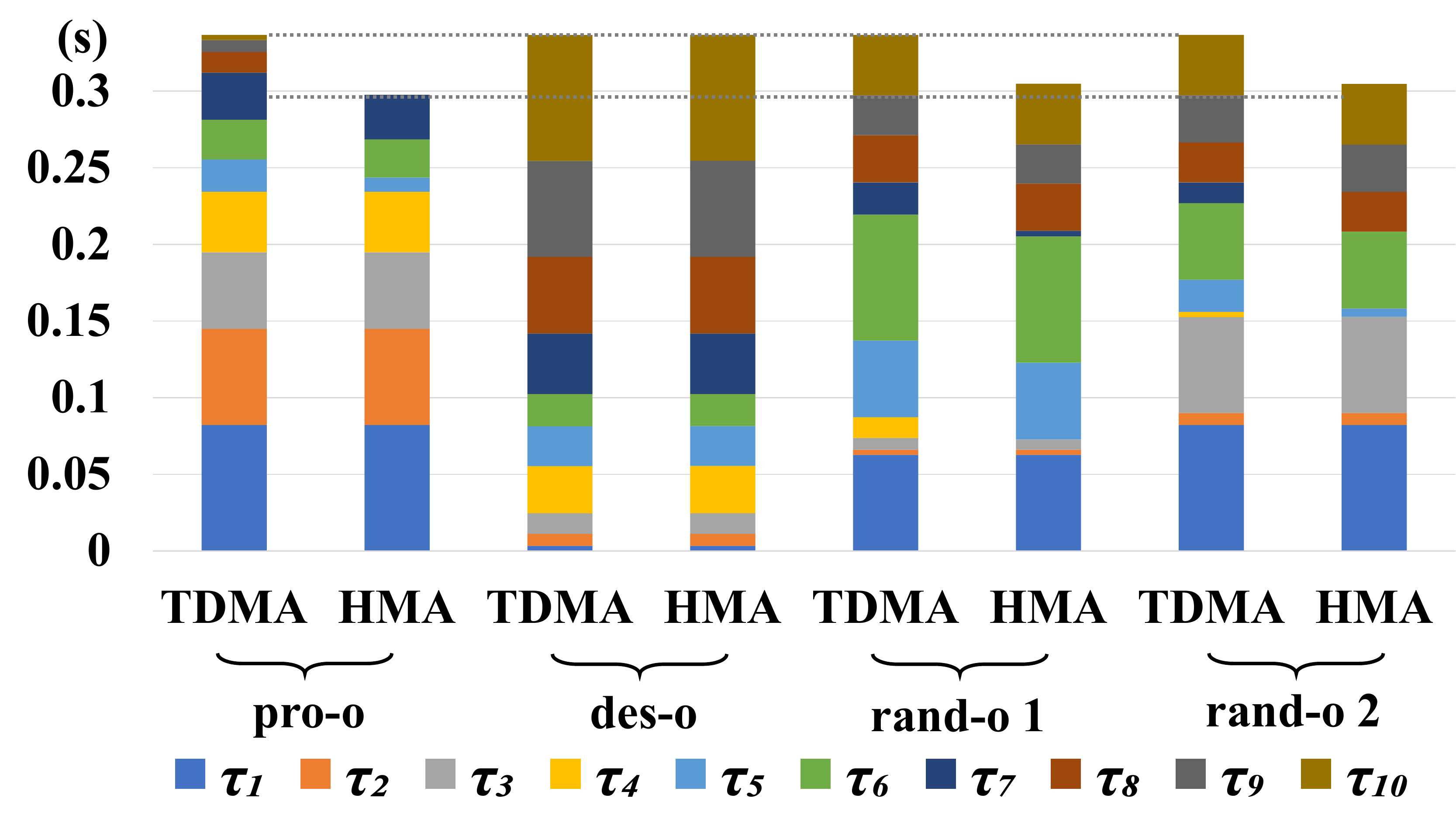}}
			\caption{The effect of different device's rearranged orders ordering ($K=10$).} 
		\label{fig:order}
	\end{figure} 

    \subsection{The Effect of Device's Ordering}
     In Fig. \ref{fig:order}, we evaluate the effect of different device's rearranged order under the transmit power constraints and transmit energy constraints, respectively. In particular, the parameters are set as $K=10$, $D_k=(5(K-k+1),0,0)$ m, $\forall k \in \mcal{K}$. Note that under this setting, there are $10!=3628800$ possible rearranged orders, which is prohibitive for utilizing the exhaustive search. Therefore, {we adopt two random  rearranged orders (\textbf{``rand-o 1''} and \textbf{``rand-o 2''}) and the descending order of TDMA-based SNR (\textbf{``des-o''}) as the benchmarks compared with our proposed ordering principle (\textbf{``pro-o''}).}
     It can be seen from Fig. \ref{fig:order} that device's order has a nonnegligible impact on the sum transmission delay when employing the proposed I-HMA transmission protocol. Specifically, {no matter under the transmit power constraints or under the transmit energy constraints, rearranging the devices according to the proposed ordering principle achieves a lower sum transmission delay than its counterparts, which validates the superiority of the proposed ordering principle.}
    
    Besides, it is observed from Fig. \ref{fig:order_p} that under the transmit power constraints, employing the proposed I-HMA transmission protocol outperforms that with I-TDMA protocol under any kind of device's orders, which validates the accuracy of Proposition \ref{pro:HMA_sup}.
    Whereas, for the transmit energy limited case as shown in Fig. \ref{fig:order_e}, this performance gain vanishes with the descending  device's order, which validates the discussions in Section \ref{sec:solu_HMA_en}1. 
    Moreover, it is noticed that employing the proposed I-HMA protocol reduces transmission completion time duration not only from the system-centric view, but also from the device-centric view, which thus can reduce the energy consumption for each device, revealing another advantage with the proposed I-HMA protocol.
    
	\section{Conclusion}
	In this paper, we investigated an IRS-aided UL transmission system, where {an} I-HMA protocol was designed to shorten the transmission delay.
    Two typical communication scenarios, namely the transmit power limited case and the transmit energy limited case were considered, where the device's rearranged order, time and power allocation, as well as dynamic IRS BF patterns were jointly optimized to minimize the sum transmission delay.
    After formulating the corresponding problems mathematically, we compared the performance of the proposed hybrid protocol with the other two conventional MA strategies through a rigorous analysis and then provided an efficient ordering principle as well as a well-designed algorithm by deeply exploiting the properties of the formulated problems.
    Through simulation results, the accuracy of the proposed theoretical analysis and the superiority of the proposed protocol were validated, which can shed some light in practice. Besides, it was found that the proposed protocol can significantly reduce the sum transmission delay by combining the additional gain of dynamic IRS BF with the time-saving advantages of the NOMA protocol, even with a small size of IRS. Furthermore, it revealed that the time reduction with the proposed design not only came from the system-centric aspect, but also from the device-centric aspect, which thus can reduce the energy consumption for each device. {The application in the downlink system will be an interesting extension and is left for future study.}
    
    \appendices
    \vspace{0.2cm}
    \section*{Appendix A: \textsc{Proof of Proposition \ref{pro:HMA_sup}}}\label{app:pro_HMA_sup}
    \vspace{0.2cm}
    {\it Proof:}
    To proceed, we first introduce the following lemma. 
    \begin{Lem}\label{lem:tight}
    We can always construct an optimal solution for (P1) and/or (P2) by setting constraints (\ref{eq:con_Lk_P1}) to be $\sum_{i=1}^k \tau_i r_{{\pi(k)},i} = L_{\pi(k)}, \forall k \in \mcal{K}$. 
    \end{Lem}
    Lemma \ref{lem:tight} can be proved as follows. If there exists a solution $\mathscr{S}^*_K({\bm{\Psi}})\teq \{{\bm{\tau}}^*, {\bm{p}}^* | \hspace{0.05cm}{\bm{\Psi}}  \}$ that achieves the minimum sum transmission delay where the constraints (\ref{eq:con_Lk_P1}) are not met with equality for some devices, i.e., $\exists \hspace{0.1cm} {\pi(k)} \in \mcal{K}$ such that $\sum_{i=1}^k \tau_i^* r_{{\pi(k)},i}({\bm{p}}^* ,{\bm{\Psi}}) > {L}_{\pi(k)}$, we can always construct another solution $\hat {\mathscr{S}}_K ({\bm{\Psi}})\teq \{\hat {\bm{\tau}}, \hat{\bm{p}} |\hspace{0.05cm}{\bm{\Psi}}  \}$  by lowering its transmit power $p_{{\pi(k)},i}$, $\forall i \in \mcal{I}(k) $ until this constraint meets with equality, i.e., by setting $\hat{p}_{{\pi(k)},i}=\alpha p^*_{{\pi(k)},i}$, $\forall i \in \mcal{I}(k) $, $\hat{p}_{{\pi(j)},i}= p^*_{{\pi(j)},i}$, $\forall j\neq k,\forall i \in \mcal{I}(j)$, $ \hat {\bm{\tau}}={\bm{\tau}}^*$, where $\alpha < 1$ such that $\sum_{i=1}^k \hat{\tau_i} r_{{\pi(k)},i}(\hat{\bm{p}} ,{\bm{\Psi}}) = {L}_{\pi(k)}$. It is easy to verify that the newly constructed solution $\hat {\mathscr{S}}_K({\bm{\Psi}}) $ satisfies all the constraints in (P1) and (P2) while keeping the sum transmission delay unchanged. Besides, after reducing the transmit power of $D_{\pi(k)}$, the interference to the latter devices i.e., $D_{\pi(n)}$, $(n>k)$, can get smaller, which can shorten $D_{\pi(n)}$'s UL transmission time consumption with the original transmit power or reduce the transmit power to finish its UL transmission task within the original time duration.

    Based on the optimal solution of the problems with the I-TDMA protocol and the proposed I-HMA protocol, i.e., $\mathscr{S}^{\td ^*}({\bm{\Psi}})$ and $\mathscr{S}^{\hy ^*}({\bm{\Psi}})$, we can construct another feasible solution $\tilde{\mathscr{S}}({\bm{\Psi}})\teq \{ \tilde{\bm{\tau}}, \tilde{\bm{p}} \hspace{0.05cm} |\hspace{0.05cm} {\bm{\Psi}} \}$ for the problem with the proposed I-HMA protocol, where  $\tilde{\bm{\tau}}={\bm{\tau}}^{\td ^*}$, $\tilde{p}_{\pi(k),k} = {p}_{\pi(k)}^{\td ^*}$, $\tilde{p}_{\pi(k),i} =0$, $\forall i \in \mcal{I}(k-1) $, $\forall k \in \mathcal{I}(K-1)$, and  $\tilde{p}_{\pi(K),i} = {p}_{\pi(K)}^{\td ^*}$, $\forall i \in \mcal{I}(K) $. Then, we have
    
    \begin{small}
    \begin{equation}
    \setlength\abovedisplayskip{-0.01cm}
    \begin{aligned}
    \tau_K^{\td ^*} = \frac{{L}_{\pi(K)} }{r_{{\pi(K)},K}( {\bm{p}}^{\td ^*} ,{\bm{\Psi}})}  &\overset{\text{(a)}}{>} \frac{{L}_{\pi(K)} - \sum_{i=1}^{k-1} \tilde{\tau}_i r_{{\pi(K)},i}( \tilde{\bm{p}} ,{\bm{\Psi}})}{r_{{\pi(K)},K}( \tilde{\bm{p}} ,{\bm{\Psi}})}\\
    &\overset{\text{(b)}}{\geq} \frac{{L}_{\pi(K)} - \sum_{i=1}^{K-1} \tau_i^{\hy ^*} r_{{\pi(K)},i}({\bm{p}}^{\hy ^*} ,{\bm{\Psi}} )}{r_{{\pi(K)},K}({\bm{p}}^{\hy ^*} ,{\bm{\Psi}} )}  \overset{\text{(c)}}{=} \tau_K^{\hy ^*}, 
    \end{aligned}
    \end{equation}
    \end{small}

    \noindent where the strict inequality (a) holds due to the fact that $\sum_{i=1}^{K-1} \tilde{\tau}_i r_{{\pi(K)},i}(${\small$\tilde{\bm{p}} ,{\bm{\Psi}}$}$)>0$ and $r_{{\pi(K)},K}( ${\small${\bm{p}}^{\td ^*} ,{\bm{\Psi}}$}$)$ $=r_{{\pi(K)},K}( \tilde{\bm{p}} ,{\bm{\Psi}})$. The inequality (b) holds since given the IRS BF patterns set ${\bm{\Psi}}$, the constructed solution $\tilde{\mathscr{S}}({\bm{\Psi}})$ is only a feasible solution of (P1), whereas the solution $\mathscr{S}^{\hy ^*}({\bm{\Psi}})$ is the optimal solution of (P1). The equality (c) holds due to the fact that we can always construct a solution by tightening the constraints (\ref{eq:con_Lk_P1}) in (P1) that achieve the minimum sum transmission delay, as indicated by Lemma \ref{lem:tight}. Therefore, we have $\mathsf{T_{sum}^{hy}}({\bm{\Psi}}) < \mathsf{T_{sum}^{td}}({\bm{\Psi}})$. 
    On the other hand, the minimum sum transmission delay with the proposed I-HMA protocol and with the I-TDMA protocol, i.e., $\mathsf{T_{sum}^{hy^*}}$ and $\mathsf{T_{sum}^{td^*}}$, are respectively achieved at $\mathscr{S}^{\hy ^*}({\bm{\Psi}}^{\hy ^*})$ and $\mathscr{S}^{\td ^*}({\bm{\Psi}}^{\td ^*})$, which leads to $\mathsf{T_{sum}^{hy}}({\bm{\Psi}}^{\hy^*})<\mathsf{T_{sum}^{hy}}({\bm{\Psi}}^{\td^*})$.\hfill$\blacksquare$
    
    \section*{Appendix B: Proof of Proposition \ref{pro:HMA_eq_NOMA}}\label{app:HMA_eq_NOMA}
     \vspace{0.2cm}
    \begin{myproof}
    When ${L}_{\pi(k)} \leq   {L}_{\pi(k)}^{{\no}}( \tv_1) ,\forall k>1$, with ${L}_{\pi(k)}^{{\no}}( \tv_1)$ given by (\ref{eq:L_bound}), it follows that 
    \begin{subequations}
    \begin{numcases}{}
     \tau_{1}^{\no ^*}\log_2 [1+P_{\pi(1)}^{\m}\gamma_1(\tv_1)] = \bar{L}_{\pi(1)}, \\
     \tau_1^{\no ^*}  \log_2 \left[1+\frac{P_{{\pi(k)}}^{\m}{\gamma}_k(\tv_1)}{1+\sum_{i=1}^{k-1} P_{{\pi(i)}}^{\m}{\gamma}_i(\tv_1)} \right]  \geq  \bar{L}_{\pi(k)}, k>1.
    \end{numcases}
    \end{subequations}
    
    \noindent It can be easily verified that $\mathsf{T_{sum}^{hy}}( {\bm{\Psi}}) = \mathsf{T_{sum}^{no}}( {\bm{\Psi}})= \tau_{1}^{\no ^*}$, which thus completes the proof.
    \end{myproof}
    
    \section*{Appendix C: Proof of Proposition \ref{pro:HMA_sup_no}}\label{app:HMA_sup_no}
    \vspace{0.2cm}
    {\it Proof:}
    If ${L}_{\pi(2)} >   {L}_{\pi(2)}^{\no}( \tv_1)$, then for the  I-NOMA protocol, it follows that
    \begin{subequations}
    \begin{numcases}{}
    p_{\pi(1),1}^{\no ^*} < P_{\pi(1)}^{\m}, p_{{\pi(2)},1}^{\no ^*}=P_{\pi(2)}^{\m},\label{con:NOMA1}\\
     \frac{\log_2 \left[1+\frac{p_{{\pi(2)},1}^{\no ^*}{\gamma}_2(\tv_1)}{1+ p_{{\pi(1)},1}^{\no ^*}{\gamma}_1(\tv_1)} \right]}{\log_2 [1+p_{\pi(1),1}^{\no ^*}\gamma_1(\tv_1)]}    =  {{L}_{\pi(2)}}\big/\hspace{-0.02cm}{{L}_{\pi(1)}},
    \end{numcases}\label{eq:NOMA_K2}
    \end{subequations}
    
    \noindent In contrast, for the proposed I-HMA protocol, we have $p_{{\pi(2)},1}^{\hy ^*}=p_{{\pi(2)},2}^{\hy ^*}=P_{\pi(2)}^{\m}$, and
   
   {\setlength\abovedisplayskip{-0.1cm}
        \begin{align}
        \mathsf{T_{sum}^{hy}}\left(p_{\pi(1),1}, {\bm{\Psi}}\right)&=\tau_{1}+\tau_{2}= \mathscr{T} \left(p_{\pi(1),1}, {\bm{\Psi}}\right)+\frac{\bar{L}_{\pi(1)}+\bar{L}_{\pi(2)}}{\log_2 [1+p_{\pi(2),2}\gamma_2(\tv_2)]},
        \end{align}}

        \noindent where $\mathscr{T} (p_{\pi(1),1},{\bm{\Psi}})$ and its partial derivative w.r.t. $p_{\pi(1),1}$ are respectively given by 
        {\setlength\belowdisplayskip{-0.2cm}
        \begin{equation}
        \begin{aligned}
        & \hspace{0.4cm} \mathscr{T}(p_{\pi(1),1},{\bm{\Psi}} )\teq\frac{\bar{L}_{\pi(1)}}{\log_2 [1+p_{\pi(1),1} \gamma_1(\tv_1 )]}\bigg\{1+\frac{\log_2 [1+p_{\pi(1),1} +p_{{\pi(2)},1} {\gamma}_2(\tv_1 )}{\log_2 [1+p_{\pi(2),2} \gamma_2(\tv_2 )]}\bigg\}.
        \end{aligned}\label{eq:T_p}
        \end{equation}}
        \begin{align}
        \frac{\partial \hspace{0.03cm}\mathscr{T}(p_{\pi(1),1},{\bm{\Psi}} )}{\partial\hspace{0.03cm} p_{\pi(1),1} }=\hspace{0.03cm}& \frac{\gamma_1(\tv_1 )\bar{L}_{\pi(1)}}{\log_2 [1+p_{\pi(1),1} \gamma_1(\tv_1 )]}\bigg\{\frac{1}{\log_2 [1+p_{\pi(2),2} \gamma_2(\tv_2 )]}\frac{1}{1+p_{\pi(1),1} +p_{{\pi(2)},1} {\gamma}_2(\tv_1 )} \notag \\
        &-\frac{1}{\log_2 [1+p_{\pi(2),2} \gamma_2(\tv_2 )]}\frac{1}{1+p_{\pi(1),1} } \frac{\log_2 [1+p_{\pi(1),1} +p_{{\pi(2)},1} {\gamma}_2(\tv_1 )}{\log_2 [1+p_{\pi(1),1} \gamma_1(\tv_1 )]} \notag \\
         &-\frac{1}{\log_2 [1+p_{\pi(1),1} \gamma_1(\tv_1 )]}\frac{1}{1+p_{\pi(1),1} }\bigg\} <0.\label{eq:T_p_d}
        \end{align}

         \noindent It can be seen from (\ref{eq:T_p_d}) that $\mathsf{T_{sum}^{hy}}\left(p_{\pi(1),1}, {\bm{\Psi}}\right)$ is monotonically decreasing as $p_{\pi(1),1}$ increases, whose minimum value is achieved at $p_{\pi(1),1}^{\hy ^*}=P_{\pi(1)}^{\m}$. On the other hand, as indicated by (\ref{eq:NOMA_K2}), there is $p_{\pi(1),1}^{\no ^*} < P_{\pi(1)}^{\m}=p_{\pi(1),1}^{\hy ^*}$. Thus, we have $\mathsf{T_{sum}^{hy}}\left(p_{\pi(1)}^{\hy^*},{\bm{\Psi}}\right) < \mathsf{T_{sum}^{hy}}\left(p_{\pi(1),1}^{\no ^*},{\bm{\Psi}}\right)$. Besides, by examining the solution for the optimization problem, it is easy to verify that $ \mathsf{T_{sum}^{hy}}\left(p_{\pi(1),1}^{\no ^*},{\bm{\Psi}}\right)=\mathsf{T_{sum}^{no}}\left(p_{\pi(1),1}^{\no ^*},{\bm{\Psi}}\right)$. In addition, the minimum sum transmission delay with the proposed I-HMA protocol and with the I-NOMA protocol, i.e., $\mathsf{T_{sum}^{hy^*}}$ and $\mathsf{T_{sum}^{no^*}}$, are respectively achieved at $\mathscr{S}^{\hy ^*}({\bm{\Psi}}^{\hy ^*})$ and $\mathscr{S}^{\no ^*}({\bm{\Psi}}^{\no ^*})$, which leads to $\mathsf{T_{sum}^{hy}}({\bm{\Psi}}^{\hy^*})<\mathsf{T_{sum}^{hy}}({\bm{\Psi}}^{\no^*})$.\hfill$\blacksquare$

    \vspace{0.2cm}
    \section*{Appendix D: Proof of Proposition \ref{pro:bound}}\label{app:bound}
  \vspace{0.2cm}
    \begin{myproof}
    Since given the IRS BF patterns set ${\bm{\Psi}}$, the optimal solution of the problem with the  I-NOMA protocol, i.e., $\mathscr{S}^{\no ^*}({\bm{\Psi}})$ is always a feasible solution for the problem with the proposed I-HMA protocol, then we have $\mathsf{T_{sum}^{hy}}({\bm{\Psi}}) \leq   \mathsf{T_{sum}^{no}}({\bm{\Psi}})$, where the equality holds when $\mathscr{S}^{\no ^*}({\bm{\Psi}})$ is also the optimal solution of the problem with the proposed I-HMA protocol. To guarantee this condition, the initial energy $E_{\pi(k)}$ at the device $D_{\pi(k)}$, $\forall k>1$, needs to satisfy
    \begin{subequations}
    \begin{numcases}{}
     \tau_1 \left[1+\frac{p_{{\pi(k)},1}{\gamma}_k(\tv_1)}{1+\sum_{i=1}^{k-1} p_{{\pi(k)},1}{\gamma}_i(\tv_1)} \right]  = \bar{L}_{\pi(k)},\\
    E_{\pi(k)}\geq\tau_1 p_{{\pi(k)},1}  \teq E^{{\no}}_{\pi(k)},
    \end{numcases}\label{eq:NOMA_t1}
    \end{subequations}
   \noindent where $\tau_1$ can be obtained by solving
    \begin{subequations}
    \begin{numcases}{}
    \tau_1 \log_2 \left[1+{p_{{\pi(1)},1}{\gamma}_1(\tv_1)} \right]  = \bar{L}_{\pi(1)},\\
    \tau_1 p_{{\pi(k)},1} = E_{\pi(1)}  ,
    \end{numcases}
    \end{subequations}
    
   \noindent which leads to (\ref{eq:E_no}).
    Next, we investigate the condition when the optimal solution of the problem with the  I-TDMA protocol, i.e., $\mathscr{S}^{\td ^*}({\bm{\Psi}})$ is the optimal solution of the problem with the proposed I-HMA protocol.
    By utilizing Lemma \ref{lem:tight}, we have
    \begin{subequations}
    \begin{numcases}{}
     \tau_1^{\hy} = {{L}_{\pi(1)}}/{r_{{\pi(1)},1}( {\bm{p}}^{\hy} ,{\bm{\Psi}})},\\
     \tau_k^{\hy} = \frac{{L}_{\pi(k)} - \sum_{i=1}^{k-1} {\tau}_i^{\hy} r_{{\pi(k)},i}( {\bm{p}}^{\hy} ,{\bm{\Psi}})}{r_{{\pi(k)},k}( {\bm{p}}^{\hy} ,{\bm{\Psi}})}, \hspace{0.2cm} \forall k>1.
    \end{numcases}
    \end{subequations}
    \noindent When the proposed I-HMA protocol reduces to the  I-TDMA protocol, the optimal protocol for each device $D_{\pi(k)}$, $\forall k >1$, is to keep inactivated during the former time duration of $\sum_{i=1}^{k-1} \tau_{i}^{\hy}$ so as to preserve its stored energy to support its UL transmission only within $\tau_k^{\hy}$. Thus, $\tau_k^{\hy}$ is determined by $D_{\pi(k)}$ whereas $\tau_{i}^{\hy}$, $\forall i \in \mcal{I}(k-1) $, can be considered as a constant when designing the optimal resource allocation strategy for $D_{\pi(k)}$. As such, the sum transmission delay minimization problem can be solved by investigating a series of subproblems for each device in a successive manner, i.e., minimizing $\tau_k^{\hy}$, $\forall k \in \mcal{K}$.
    To this end, we utilize the Dinkelbach method and introduce a set of auxiliary variables ${\bm{\lambda}} \teq \{\lambda_1,\cdots,\lambda_K\}$, which satisfies
    \begin{equation}
    \begin{aligned}
    \lambda_k &= \frac{1}{\tau_k^{\hy}} = \frac{r_{{\pi(k)},k}( {\bm{p}}^{\hy} , {\bm{\Psi}})}{{L}_{\pi(k)} - \sum_{i=1}^{k-1} {\tau}_i^{\hy} r_{{\pi(k)},i}( {\bm{p}}^{\hy} , {\bm{\Psi}})}. 
    \end{aligned}
    \end{equation}
    
    \noindent Then, the optimal resource allocation solution for $D_{\pi(k)}$ can be obtained by solving
    \begin{subequations}\label{eq:Pro_KKT}
    \begin{align}
    \hspace{-1cm}\underset{ {{\lambda_k}}, {\bm{p}}^{\hy}  }{\max} \hspace{0.4cm} &    r_{{\pi(k)},k}( {\bm{p}}^{\hy} , {\bm{\Psi}})- \lambda_k  {L}_{\pi(k)} +\lambda_k \sum\nolimits_{i=1}^{k-1} {\tau}_i^{\hy} r_{{\pi(k)},i}( {\bm{p}}^{\hy} , {\bm{\Psi}})  \\
    \hspace{-0.2cm}\text { s.t. } \hspace{0.3cm}   
    &(\ref{eq:P2_con3}), \quad \sum\nolimits_{i=1}^k {p_{{\pi(k)},i}^{\hy}}/{\lambda_k}  \leq  E_{\pi(k)}, 
    \end{align}
    \end{subequations}
    \noindent which is a convex problem with the given $\{\tau_j^{\hy} \hspace{0.05cm} | \hspace{0.1cm} \forall j \in \mcal{I}(k-1)\}$, $ \{ p_{\pi(j),i}^{\hy} \hspace{0.05cm} | \hspace{0.1cm} \forall i \in \mcal{I}(j) , \forall j \in \mcal{I}(k-1)\}$, ${{\lambda_k}}$ and $ {\bm{\Psi}}$. Therefore, the optimal power allocation solution can be expressed as a function w.r.t. $\lambda_k$ by using the Karush-Kuhn-Tucker (KKT) condition \cite{ding2018delay}. By investigating the dual function of problem (\ref{eq:Pro_KKT}), if the optimal solution of $D_{\pi(k)}$ satisfies $ p_{\pi(k),i}^{\hy}=0$, $  \forall i \in \mcal{I}(k-1)$, the following inequalities must satisfy
    \begin{equation}
    \begin{aligned}
    {\gamma_{k}(\tv_k )}\big/\left({1+p_{\pi(k),k}^{\hy}}\right) \geq {\gamma_{k}(\tv_i )}\big/\left({1+p_{\pi(i),i}^{\td}}\right), \hspace{0.2cm} \forall i \in \mcal{I}(k-1),
    \end{aligned}
    \end{equation}\noindent which further leads to (\ref{eq:E_td}) and (\ref{eq:E_td_i}). Besides, $\tau_k^{\td^*}$ and $p_{{\pi(k)}}^{\td^*}$, $\forall k\in\mcal{K}$, can be obtained by solving
    \begin{subequations}
    \begin{numcases}{}
    \tau_k \log_2 \left[1+{p_{{\pi(k)}}{\gamma}_k(\tv_k)} \right]  = \bar{L}_{\pi(k)},\\
    \tau_k p_{{\pi(k)}} = E_{\pi(k)}  ,
    \end{numcases}
    \end{subequations}
   \noindent which leads to $\tau_k^{\td^*}=\mathscr{G}_k(\tv_k)$, with $\mathscr{G}_k(\tv_k)$ given by (\ref{eq:TDMA_E}).
    \end{myproof}
    
    \vspace{0.4cm}
    \linespread{1.7}
    \bibliographystyle{IEEEtran}
    \bibliography{IEEEabrv,piaobib}
\end{document}